\documentclass{PoS}
\usepackage{epsfig}
\usepackage{amsmath}

\newcommand{\cO}{{\cal O}}
\newcommand{\Eq}[1]{Eq.~(\ref{#1})}
\newcommand{\Cone}{C_1}
\newcommand{\Ctwo}{C_2}
\newcommand{\cQ}{{\cal Q}}
\newcommand{\cR}{{\cal R}}
\newcommand{\MSbar}{{\overline{\rm MS}}}

\newcommand{\bi}{\begin{itemize}}
\newcommand{\ei}{\end{itemize}}
\newcommand{\ben}{\begin{enumerate}}
\newcommand{\een}{\end{enumerate}}
\newcommand{\beq}{\begin{equation}}
\newcommand{\eeq}{\end{equation}}
\newcommand{\beqn}{\begin{eqnarray}}
\newcommand{\eeqn}{\end{eqnarray}}
\newcommand{\bean}{\begin{eqnarray*}}
\newcommand{\eean}{\end{eqnarray*}}
\newcommand{\bea}{\begin{eqnarray}}
\newcommand{\eea}{\end{eqnarray}}
\newcommand{\nn}{\nonumber}

\newcommand{\ri}{{\rm i}}
\newcommand{\bx}{\mathbf{x}}

\title{First quenched results for the matrix elements of the $B_{B_s}$
 mixing parameter in the static limit from tmQCD\thanks{CERN-PH-TH-2007-175, FTUAM-07-15, 
IFT-UAM-CSIC-07-46, MKPH-T-07-11}}

\ShortTitle{Preliminary results of $B_{B_s}$ in the static limit from tmQCD }

\author{\epsfig{figure=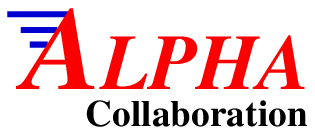,width=2.5cm}}

\author{\speaker{Filippo Palombi and Mauro Papinutto}\\
  CERN, Physics Department, Theory Division, CH-1211 Geneva 23,
  Switzerland\\
  E-mail: \email{filippo.palombi@cern.ch,mauro.papinutto@cern.ch}}
\author{Carlos Pena\\
         Departamento de F\'{\i}sica Te\'orica C-XI and
         Instituto de F\'{\i}sica Te\'orica C-XVI\\
                Facultad de Ciencias, Universidad Aut\'onoma de Madrid,
                Cantoblanco E-28049 Madrid, Spain\\
                E-mail: \email{carlos.pena@uam.es}}
\author{
  Hartmut Wittig\\
  Institut f\"ur Kernphysik, University of Mainz, D-55099 Mainz, Germany\\
  E-mail: \email{wittig@kph.uni-mainz.de}}

\abstract{We report on a non-perturbative study of the scale-dependent 
renormalization factors of a multiplicatively renormalizable basis of 
$\Delta B=2$ parity-odd four-fermion operators in quenched lattice QCD. 
We also present some preliminary results of the matrix elements related 
to the mixing parameter of the $B_s$-meson. In our lattice formulation, the 
heavy quark is treated in the static approximation, while the strange one 
belongs to a doublet of twisted mass fermions at full twist, i.e. with twist 
angle $\alpha=\pi/2$. In this framework, the parity-even $\Delta B=2$ four-fermion 
operators responsible for the mixing are rotated onto a linear combination of 
parity-odd operators in the above-mentioned basis. Their physical matrix elements 
between static $B_s$-mesons are extracted from lattice correlators with 
Schr\"odinger functional boundary conditions. We observe a suppression of excited 
state contributions to the $B_{B_s}$ mixing parameter and speculate about 
possible explanations.}

\FullConference{The XXV International Symposium on Lattice Field Theory\\
                July 30-4 August 2007\\
                Regensburg, Germany}

\begin{document}

\section{Introduction}

Currently, theoretical determinations of the mixing parameters $B_{B_q}$ ($q=d,s$)
are becoming more and more urgent in relation to the Unitarity
Triangle analysis. The $B$-parameters are defined as the relative deviations 
from the Vacuum Saturation Approximation (VSA) of the matrix elements 
of $\Delta B=2$ four-fermion operators between $B$-meson states, i.e.
\begin{equation}
\label{Bpars}
B_{B_q} = \frac{\langle \bar B^0_q|
\cO^{\scriptscriptstyle\Delta B=2}_{\rm\scriptscriptstyle LL}
 |B^0_q\rangle}{ \frac{8}{3} f_{B_q}^2m_{B_q}^2}\ , \qquad 
\cO^{\scriptscriptstyle\Delta B=2}_{\rm\scriptscriptstyle LL} =
\left[\bar\psi_b\gamma_\mu(1-\gamma_5)\psi_q\right]\left[\bar\psi_b\gamma_\mu(1-\gamma_5)\psi_q\right]\ .
\end{equation}
They encode the low-energy information related to particle-antiparticle 
oscillations and are formally accessible to lattice QCD simulations. 
Nevertheless, a direct computation of $B_{B_q}$ is hampered by the presence
of the large value of the $b$-quark mass, which imposes the adoption of tiny
lattice spacings $\left(a\ll 1/(5{\rm GeV})\right)$ in order to avoid large lattice
artefacts. A possible way out is to expand the $B$-parameters in 
Heavy Quark Effective Theory (HQET), i.e. in inverse powers of the $b$-quark 
mass. The leading contribution, also known as the static approximation, is
expected not to be far from the relativistic value, as previous lattice results
have shown. Even so, the na\"ive
lattice discretization of the effective four-fermion operators of the static 
theory, based on Wilson-type light fermions, is affected by a non-trivial 
renormalization mixing, due to the explicit breaking of chiral symmetry, which 
pushes the numerics up to the edge of our current technology. Although 
Ginsparg-Wilson fermions appear as the natural discretization to study 
left-left four-quark operators, we follow a computationally cheaper approach, 
based on twisted mass QCD (tmQCD) \cite{Frezzotti:2000nk}, which allows for 
purely multiplicative renormalization at the same computational cost as with 
Wilson quarks.  

\section{Computational strategy}

Our starting point is the equation relating the left-left
operator of the fully relativistic theory to the four-fermion operators of 
HQET, 
\begin{equation}
\label{ptmatching}
\hskip -0.8cm\cO^{\scriptscriptstyle\Delta B=2}_{\rm\scriptscriptstyle LL}(m_b) = \ \ 
\Cone(m_b,\mu) Q_1(\mu) + \Ctwo(m_b,\mu)Q_2(\mu) + O\left(\frac{1}{m_b}\right)\ , \\[-2.5ex]
\end{equation}
\begin{alignat}{3}
\hskip 2.62cm Q_1 & \ = \ \cO^{\rm stat}_{\rm\scriptscriptstyle VV+AA} & \ = \ &\left(\bar\psi_{h}\gamma_\mu\psi_q\right)\left(\bar\psi_{\bar h}\gamma_\mu\psi_q\right) + \left(\bar\psi_{h}\gamma_\mu\gamma_5\psi_q\right)\left(\bar\psi_{\bar h}\gamma_\mu\gamma_5\psi_q\right)\ , \\[1.9ex]
Q_2 & \ = \ \cO^{\rm stat}_{\rm\scriptscriptstyle SS+PP} & \ = \ &\left(\bar\psi_{h}\psi_q\right)\left(\bar\psi_{\bar h}\psi_q\right) + \left(\bar\psi_{h}\gamma_5\psi_q\right)\left(\bar\psi_{\bar h}\gamma_5\psi_q\right)\ .
\end{alignat}
\Eq{ptmatching} has to be understood as a scheme dependent perturbative 
matching between two renormalizable field theories. The coefficients $C_i$, 
known at NLO in the $\MSbar$/NDR scheme \cite{Buchalla:1996ys}, 
provide the RG evolution from the defining scale $m_b$ of the 
effective theory down to a scale $\mu\approx 1\ {\rm GeV}$. 
Although a natural hierarchy $\mu<m_b$ has to be assumed in the matching equation, 
it should be observed that, due to the renormalizability of the static theory, 
the four-fermion operators $Q_{1,2}$ are perfectly defined at any scale. In 
particular, they can be perturbatively evolved up to the RGI point through the 
appropriate $2\times 2$ static anomalous dimension matrix, i.e.
\begin{equation}
\left[ \begin{array}{c} Q_1^{\rm\scriptscriptstyle RGI} \\ Q_2^{\rm\scriptscriptstyle RGI} \end{array} \right] 
= {\hat c}(\mu)
\left[ \begin{array}{c} Q_1(\mu) \\ Q_2(\mu) \end{array} \right] \ ,
\end{equation}
where ${\hat c}(\mu)$ will be defined later.

The advantage of introducing RGI operators is twofold. On the one hand, 
they are truly non-perturbative quantities, free of systematic uncertainties
related to perturbative truncations. On the other, they are regularization
independent. As such they can be linked to any specific lattice 
regularization, to be chosen on the basis of computational convenience. 
A simplification of the renormalization pattern is achieved if we perform 
a change of basis, i.e. we introduce the primed operators
\begin{equation}
\left[ \begin{array}{c} Q_1'^{\rm\scriptscriptstyle RGI} \\ 
                        Q_2'^{\rm\scriptscriptstyle RGI} \end{array} \right] = 
\left[ \begin{array}{c} Q_1^{\rm\scriptscriptstyle RGI} \\
                        Q_1^{\rm\scriptscriptstyle RGI} + 4Q_2^{\rm\scriptscriptstyle RGI} \end{array} \right] =
\left[ \begin{array}{cc} 1 & 0 \\ 1 & 4 \end{array} \right] \left[ \begin{array}{c} Q_1^{\rm\scriptscriptstyle RGI} \\
                        Q_2^{\rm\scriptscriptstyle RGI} \end{array} \right] = 
                        \cR \left[ \begin{array}{c} Q_1^{\rm\scriptscriptstyle RGI} \\
                                                    Q_2^{\rm\scriptscriptstyle RGI} \end{array} \right]\ .
\end{equation}
This redefinition becomes particularly advantageous on the lattice if the relativistic degrees of freedom
are discretized according to tmQCD at full twist, i.e. with twist angle $\alpha=\pi/2$ \cite{Frezzotti:2000nk}. 
In particular, from now on we consider the specific case of the $B_s$-meson, for which we assume a fermion content 
made of a static quark plus a twisted strange quark belonging to a fully twisted $(c,s)$-doublet. 
Lighter degrees of freedom, i.e. the up and down quarks, do not need to be further specified, 
since they do not enter the valence sector\footnote{This freedom 
allows to extend the present strategy to $N_{\rm f}=2$ with any kind of dynamical sea, without incurring in 
mixed action issues, such as the adoption of different lattice regularizations for valence and sea 
quarks.}. In tmQCD the operators $Q_{1,2}'$ are mapped onto their
odd parity counterparts $\cQ_{1,2}'$, which renormalize purely multiplicatively, as proved in 
\cite{Palombi:2006pu}. In other words, with some abuse of
notation 
\begin{align}
\langle Q_1'^{\rm RGI}\rangle & = \lim_{a\to 0} {\hat Z}'_{1,\rm RGI}\left(g_0(a)\right)\langle\cQ_1'(a)\rangle_{\rm tmQCD}^{\alpha=\pi/2}\ , \nonumber \\
\langle Q_2'^{\rm RGI}\rangle & = \lim_{a\to 0} {\hat Z}'_{2,\rm RGI}\left(g_0(a)\right)\langle\cQ_2'(a)\rangle_{\rm tmQCD}^{\alpha=\pi/2}\ ,
\end{align}
where
\begin{alignat}{3}
\hskip -4.0cm \cQ'_1 & \ = \ \cO_{\rm\scriptscriptstyle VA+AV} & \ = \ &\left(\bar\psi_{h}\gamma_\mu\psi_q\right)\left(\bar\psi_{\bar h}\gamma_\mu\gamma_5\psi_q\right) + \left(\bar\psi_{h}\gamma_\mu\gamma_5\psi_q\right)\left(\bar\psi_{\bar h}\gamma_\mu\psi_q\right)\ , \\[1.9ex]
\cQ'_2 & \ = \ \cO_{\rm\scriptscriptstyle VA+AV} + 4\cO_{\rm\scriptscriptstyle PS+SP} & \ = \ &\left(\bar\psi_{h}\gamma_\mu\psi_q\right)\left(\bar\psi_{\bar h}\gamma_\mu\gamma_5\psi_q\right) + \left(\bar\psi_{h}\gamma_\mu\gamma_5\psi_q\right)\left(\bar\psi_{\bar h}\gamma_\mu\psi_q\right) + \nonumber \\[1.7ex]
& & & 4\left[\left(\bar\psi_{h}\gamma_5\psi_q\right)\left(\bar\psi_{\bar h}\psi_q\right) + 
\left(\bar\psi_{h}\psi_q\right)\left(\bar\psi_{\bar h}\gamma_5\psi_q\right)\right]\ .
\end{alignat}
The RGI renormalization constants ${\hat Z}'_{k,\rm RGI}$ $(k=1,2)$ have been recently 
obtained in the quenched approximation \cite{Palombi:2007dr} through finite size 
techniques based on the Schr\"odinger functional \cite{Luscher:1992an}. Since the latter
allows for the adoption of mass independent schemes, the computation of ${\hat Z}'_{k,\rm RGI}$ has been 
performed with standard (untwisted) Wilson fermions. A preliminary study of the 
non-perturbative renormalization for $N_{\rm f}=2$ has been also presented at this conference
\cite{Pena:lat07}. 

\section{Non-perturbative renormalization in the Schr\"odinger functional} 
\label{NPrenSF}

In order to study the renormalization of the four-quark operators, we consider
a theory with a light quark sector consisting of two massless ${\rm O}(a)$ improved
Wilson-type quarks $(\psi_1,\psi_2)$ entering the four-quark operators, plus a
third light spectator quark $\psi_3$, regularized in the same way, whose r\^ole
will be clarified in a moment. Suitable renormalization conditions can be specified
in terms of SF correlators made of bilinear boundary source operators 
$\Sigma_{s_1 s_2}$, $\Sigma'_{s_1 s_2}$ (lying resp. on the two time 
boundaries $x_0=0$ and $x_0=T$)
\begin{equation}
\label{bilinearsource}
\Sigma_{s_1 s_2}[\Gamma] =
a^6\sum_{{\mathbf x}, {\mathbf y}}\bar{\zeta}_{s_1}({\mathbf x})\Gamma\zeta_{s_2}({\mathbf y})\ , \qquad\qquad
\Sigma'_{s_1s_2}[\Gamma] = a^6\sum_{{\mathbf x},{\mathbf y}}\bar{\zeta}'_{s_1}({\mathbf x})\Gamma\zeta'_{s_2}({\mathbf y}) \ ,
\end{equation}
and the four-fermion operators ${\cal Q}'_{1,2}$. Due to the flavour 
and parity structure of ${\cal Q}'_{1,2}$, zero-momentum correlators 
need at least three bilinear boundary sources. 
Two bilinears are placed at $x_0=0$ 
and the third one at $x_0=T$. Their product gives rise to a
generalized source
\begin{equation}
{\cal W}[{\Gamma}_1,{\Gamma}_2,{\Gamma}_3] =
\Sigma_{1h}[{\Gamma}_1]\Sigma_{23}[{\Gamma}_2]\Sigma'_{3\bar h}[{\Gamma}_3] \ ,
\end{equation}
which is parity-odd under five different choices of the Dirac 
matrices $\Gamma_1$, $\Gamma_2$ and $\Gamma_3$, i.e.
\bea
{\cal S}^{(1)} = {\cal W}[\gamma_5,\gamma_5,\gamma_5] \ ,\qquad\qquad
{\cal S}^{(2)} = \frac{1}{6}\sum_{k,l,m=1}^3\epsilon_{klm}{\cal W}[\gamma_k,\gamma_l,\gamma_m] \ ,\qquad \nn \\
{\cal S}^{(3)} = \frac{1}{3}\sum_{k=1}^3{\cal W}[\gamma_5,\gamma_k,\gamma_k]\ , \quad 
{\cal S}^{(4)} = \frac{1}{3}\sum_{k=1}^3{\cal W}[\gamma_k,\gamma_5,\gamma_k]\ , \quad 
{\cal S}^{(5)} = \frac{1}{3}\sum_{k=1}^3{\cal W}[\gamma_k,\gamma_k,\gamma_5]\ .
\eea
All of the above sources have the same quantum numbers as 
${\cal Q}'_{1,2}$ and can be used as probes within the correlators
\begin{equation}
\label{4Fcorr}
{\cal F}^{(s)}_k(x_0) = L^{-3}\langle {\cal Q}'_k(x) {\cal S}^{(s)}\rangle \ .
\end{equation}
Nevertheless, their renormalization is non-trivial and requires the 
introduction of multiplicative renormalization constants to 
absorb the additional logarithmic divergences of the boundary fields 
from Eq.~(\ref{4Fcorr}). To avoid this, we introduce some 
boundary-to-boundary correlators
\bea
f_1^{hl} & = & -\frac{1}{2L^6}\langle\Sigma'_{1\bar h}[\gamma_5]\Sigma_{h1}[\gamma_5]\rangle \ , \\[1.5ex]
f_1^{ll} & = & -\frac{1}{2L^6}\langle\Sigma'_{12}[\gamma_5]\Sigma_{21}[\gamma_5]\rangle \ ,  \\[1.5ex]
k_1^{ll} & = & -\frac{1}{6L^6}\sum_{k=1}^3\langle\Sigma'_{12}[\gamma_k]\Sigma_{21}[\gamma_k]\rangle\ , 
\eea
and use them in the ratios
\bea
\label{ratios}
{h_{k;\alpha}^{(s)}(x_0)} = \frac{{\cal F}_k^{(s)}(x_0)}{f_1^{hl}[f_1^{ll}]^{1/2-\alpha}[k_1^{ll}]^\alpha}\ , 
\eea
in such a way that the additional renormalization factors of the boundary sources 
in Eq.~(\ref{4Fcorr}) drop out. The parameter $\alpha$ in the exponent of $f_1^{ll}$ and $k_1^{ll}$
can be freely chosen without changing the flavour content of the denominator and, 
in what follows, it will take values $\alpha=0,1/2$. 
\vskip 0.2cm
Renormalization conditions, formulated in terms of the ratios ${h_{k;\alpha}^{(s)}(x_0)}$,
read 
\bea
{
\label{renfac}
{\cal Z}'^{(s)}_{k;\alpha}}(g_0,\mu\equiv 1/L)\ {h_{k;\alpha}^{(s)}(T/2)} = h_{k;\alpha}^{(s)}(T/2)|_{g_0=0}\ , 
\eea
where $T=L$, no background field is introduced and the SF $\theta$--angle \cite{Luscher:1996sc} is set to
$\theta=0.5$. In our simulations we adopt four different lattice discretizations of 
the heavy quark action, i.e. the standard Eichten-Hill one \cite{Eichten:1989zv} and its statistically improved
versions where the na\"ive parallel transporter is replaced by a smeared APE, HYP1 or HYP2 
gauge link \cite{Della Morte:2005yc}. However, in this talk we only report on results with the HYP2 action, 
i.e. the one with the best signal-to-noise ratio. Out of the plethora of renormalization
schemes that can be defined by Eq.~(\ref{renfac}),  we choose our preferred ones to be $(s,\alpha)=(1,0)$ 
for $k=1$ and $(s,\alpha)=(3,0)$ for $k=2$ ({\it cf} ref.~\cite{Palombi:2007dr} for further 
details), thus eliminating the indices $s$ and $\alpha$ from the notation.

\section{Renormalization group running} 
\label{RG running}

The formal solution of the Callan-Symanzik equation relates the scheme-dependent 
RG running operator $\cQ'_k(\mu)$ to the renormalization group invariant one 
$\left(\cQ'_k\right)_{\rm RGI}$
\bea
\label{CallanSymanzik}
\!\!\!\!\!\!\!\!\!\!\!\!\!\!\!\!\!\!\left(\cQ'_k\right)_{\rm RGI} = 
\cQ'_k(\mu)\left[\frac{\bar g^2(\mu)}{4\pi}
\right]^{-{{{\gamma}'_k}^{(0)}}/{2b_0}}
\exp\left\{-\int_0^{\bar g(\mu)} dg\left(\frac{{\gamma}'_k(g)}{\beta(g)} - 
\frac{{\gamma'}_k^{(0)}}{b_0g}\right)\right\}
= \cQ'_k(\mu) {\hat c}'_k(\mu)\ ,\eea
where $\bar g(\mu)$ is the scheme and scale-dependent renormalized coupling. 
Our goal is to compute ${\hat c}'_k(\mu)$ non-perturbatively. In practice 
the strategy we follow is to split perturbative and non-perturbative contributions 
at a high renormalization scale $\mu_{\rm pt}$, 
\bea
\label{eq:split}
\label{RGIone}
  \left(\cQ'_k\right)_{\rm RGI} =  {\hat c}'_{\:k}(\mu_{\rm pt})U'_k(\mu_{\rm pt},\mu_{\rm had})\cQ'_k(\mu_{\rm had})\ ,
\eea 
where $U'_k(\mu_{\rm pt},\mu_{\rm had})\equiv{\hat c}'_k(\mu_{\rm had})/{\hat c}'_k(\mu_{\rm pt})$ 
represents the evolution of the renormalized operators $\cQ'_k(\mu)$ from 
the low-energy hadronic scale $\mu_{\rm had}$ to the high-energy perturbative scale 
$\mu_{\rm pt}\gg \mu_{\rm had}$. Our first task has been to compute it non-perturbatively. Since 
it is difficult to accommodate scales which differ by orders of magnitude in a 
single lattice calculation, it is useful to factorize the evolution and adopt a 
recursive approach. Accordingly, we introduce the so-called step-scaling functions 
(SSFs) $\sigma_k$ and $\sigma$, which describe the change in the renormalization
constants and the gauge coupling respectively, when the energy scale $\mu$ is decreased 
by a factor~of two,
\bea
\sigma(u) = \bar{g}^2(\mu/2)\ ,\qquad\qquad\qquad\qquad\qquad u\equiv\bar{g}^2(\mu)\ ,\qquad\qquad \nn \\[1.5ex]
\sigma_k(u) = U'_k(\mu,\mu/2)^{-1}=\lim_{a \to 0}
  \frac{{\cal Z}'_k(g_0,a\mu/2)}{{\cal Z}'_k(g_0,a\mu)}
  \Big|_{u\equiv\bar{g}^2(\mu)}^{m=0}\equiv \lim_{a \to 0}\Sigma_k(u,a\mu)\ ,
\eea
and $g_0$ denotes the bare coupling. Having computed the SSFs for a sequence of 
couplings~$u_i,\,i=0,1,2,\ldots , n-1$, we can construct the non-perturbative 
evolution $U'_k(2^n\mu_{\rm had},\mu_{\rm had})$ from the product of SSFs
\bea
  U'_k(2^n\mu_{\rm had},\mu_{\rm had}) = \left\{\prod_{i=0}^{n-1}
  {\sigma}_k(u_i)\right\}^{-1},
  \quad u_i=\bar{g}^2(2^{(i+1)}\mu_{\rm had})\ .
\eea
In the present computation $\mu_{\rm had}$ is taken to be a few hundreds of MeV and 
we have chosen $n=8$, so that we could trace the evolution non-perturbatively 
over three orders of magnitude. In this way $\mu_{\rm pt}\equiv 2^n\mu_{\rm had}$ 
is large enough to allow for a perturbative evaluation of ${\hat c}'_{\:k}(\mu_{\rm pt})$ 
with the operator anomalous dimension approximated at NLO~\cite{Palombi:2006pu}) and the
$\beta$-function at NNLO~\cite{Bode:1999sm}. The relation between the RGI operators and the bare 
lattice ones defines the total RGI renormalization factor
\begin{equation}
\label{RGItwo}
  \left(\cQ'_k\right)_{\rm RGI} = {\hat Z}'_{k,\rm RGI}(g_0)
  \cQ'_k(a)\ .
\end{equation}
A comparison between Eq.~(\ref{RGIone}) and Eq.~(\ref{RGItwo}) leads to
\begin{equation}
\label{eq:totalren}
  {\hat Z}'_{k,\rm RGI}(g_0)={\hat c}'_k(\mu_{\rm pt})
    U'_k(\mu_{\rm pt},\mu_{\rm had})
    {\cal Z}'_k(g_0,a\mu_{\rm had})\ .
\end{equation}
The factor ${\cal Z}'_k(g_0,a\mu_{\rm had})$ must be determined for
each operator in a lattice simulation at fixed $\mu_{\rm had}$ for a
range of bare couplings, using suitable renormalization conditions. 
%We note that all reference to the
%scales $\mu_{\rm pt}$ and $\mu_{\rm had}$ drops out in the total
%renormalization factor.
In our simulations we have $\mu_{\rm had}=1/(2L_{\rm max})\approx 270$ MeV 
where $L_{\rm max}$ is fixed through the condition 
$\bar g^2_{\rm\scriptscriptstyle SF}(1/L_{\rm max})=3.480$. This corresponds to 
having $L_{\rm max}/r_0=0.718(16)$ ($r_0=0.5$\,fm) \cite{Necco:2001gh}. The sequence of 
couplings $u_i=\bar g^2_{\rm\scriptscriptstyle SF}(2^{-i}L_{\rm max})$ is obtained 
by solving the recursion relation 
$u_0 = 3.480$, $\sigma(u_{l+1})=u_l$, $l=0,1,\ldots$~.
 
\section{Continuum extrapolation of the step scaling functions} 

\begin{figure}[!ht]
  \begin{center}
      \hskip -0.3cm\includegraphics[width=7.0 true cm,angle=0]{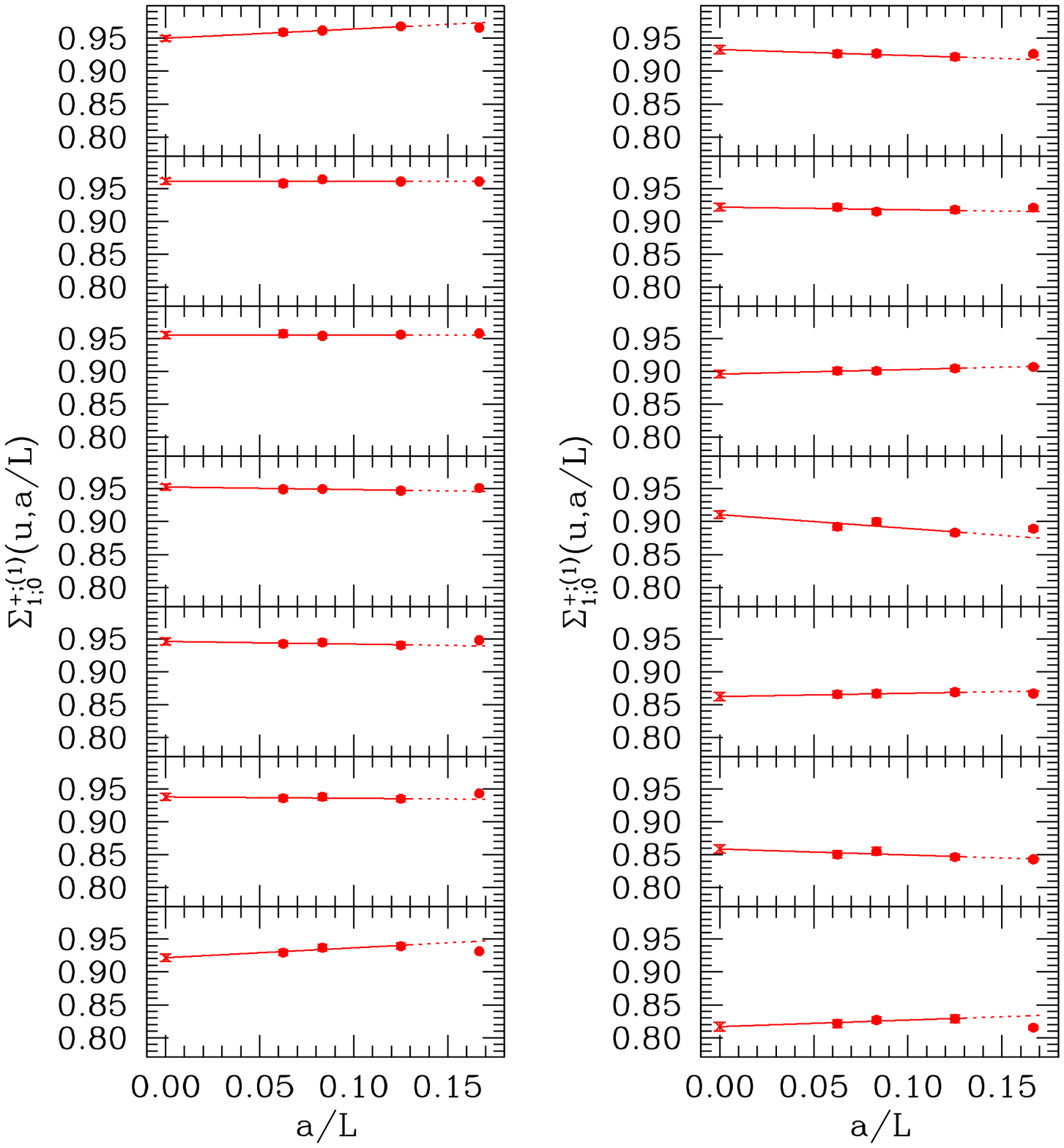}
      \hskip 0.8cm \includegraphics[width=7.0 true cm,angle=0]{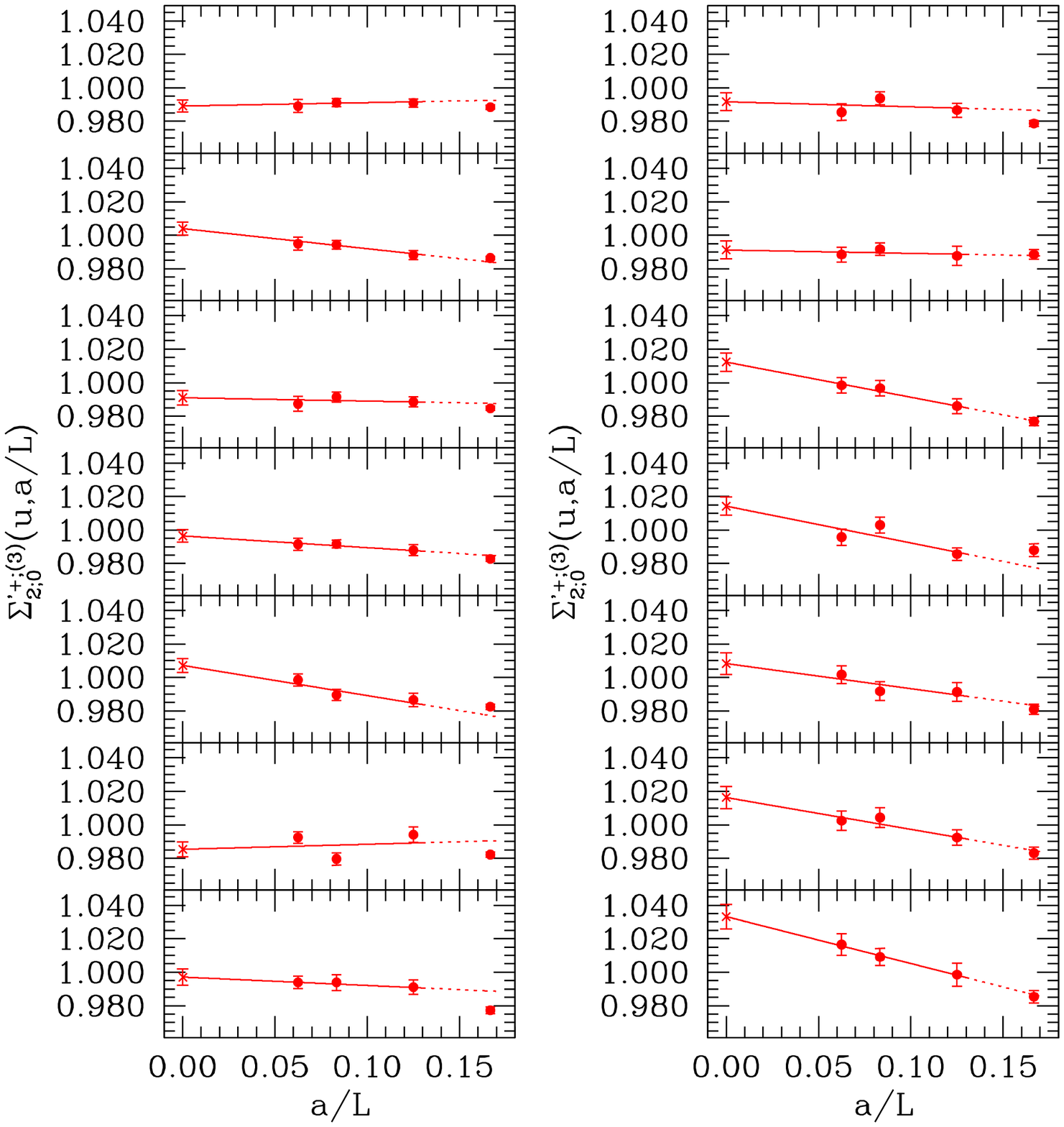}
    \caption{Continuum limit extrapolation of the SSFs
      $\Sigma_{1}$ (with $s=1,\,\alpha=0$) and  $\Sigma_{2}$ (with $s=3,\,\alpha=0$) 
      at various SF renormalized couplings computed using the HYP2 lattice discretisations 
      of the static action. The SF coupling $u$ increases from top-left to bottom-right.}
    \label{fig:extrap1}
  \end{center}
\vskip -0.6cm
\end{figure}

The lattice SSFs $\Sigma_{k}$ must be extrapolated to the
continuum limit (i.e. vanishing $a/L$) at fixed gauge coupling in
order to obtain their continuum counterparts $\sigma_{k}$. Since
the four-fermion operators have not been improved, we expect the
dominant discretization effects to be O$(a)$; therefore our data should
exhibit a linear behaviour in $a/L$. Accordingly, we have fitted to 
the {\it ansatz}
\beq
\Sigma_{k}(u,a/L) = \sigma_{k}(u) + \rho(u)\,(a/L) \,.
\eeq
Fits have been performed using either four values of the lattice 
spacing, i.e. $L/a=6,8,12,16$ or, alternatively, without taking into 
account the coarsest data $L/a=6$, which may be subject to higher-order 
lattice artefacts. The results from three- and four-point fits are
always compatible within one standard deviation for all operators and
schemes, save for a few exceptions in which the agreement drops at the
level of 1.5 standard deviations only. We have therefore decided to choose the 
three-point based linear extrapolations to extract our final estimates of 
$\sigma_{k}$. The resulting continuum limit extrapolations for our preferred 
renormalization schemes are illustrated in Fig.~\ref{fig:extrap1}.
The maximal statistical uncertainty for $\sigma_k$ is $\sim 1.5\%$ and is found
at the largest value of $u$ when discarding data at $L/a=6$. The values of $\rho$ 
obtained in the fits of $\cQ'_1$ are always compatible with zero within 
the statistical uncertainty, while in the case of $\cQ'_2$ they are not 
compatible with zero for $u\gtrsim 2$, thus signalling a stronger dependence upon 
the cut-off.  

\section{Non-perturbative RG running in the continuum limit}

\begin{figure}[!ht]
  \begin{center}
      \hskip -1.1cm\includegraphics[width=6.8 true cm,angle=0]{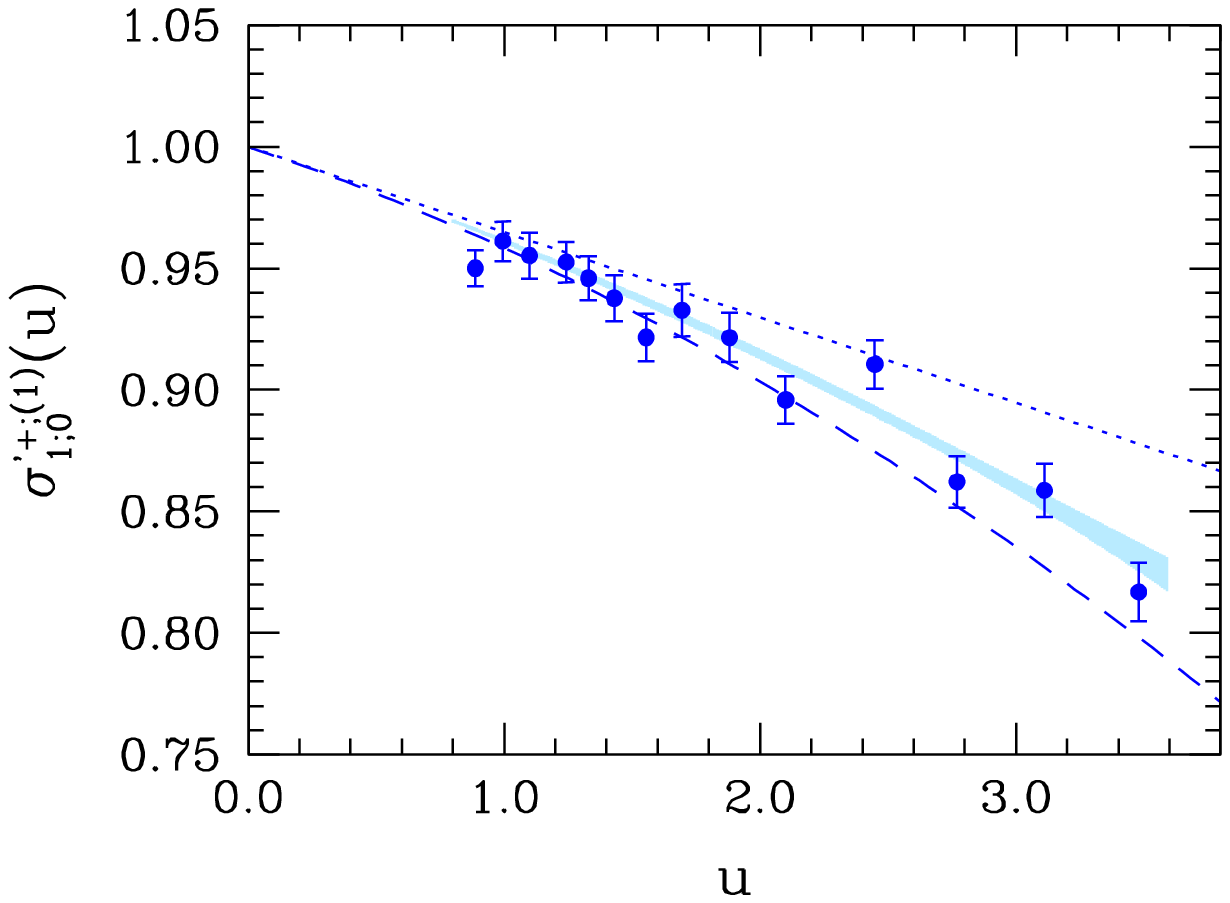}
      \hskip 0.5cm \includegraphics[width=6.8 true cm,angle=0]{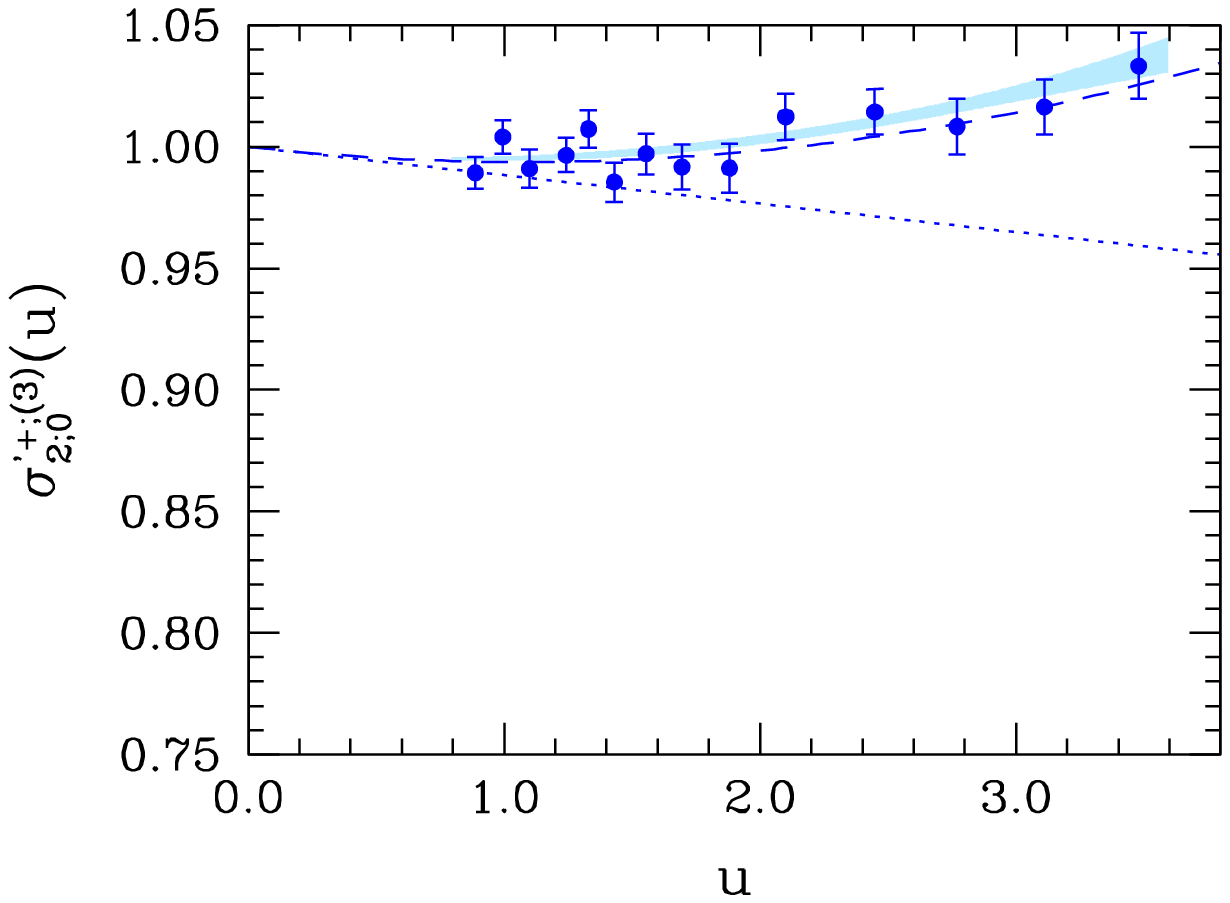}
\vskip 0.5cm
      \hskip -1.1cm\includegraphics[width=6.8 true cm,angle=0]{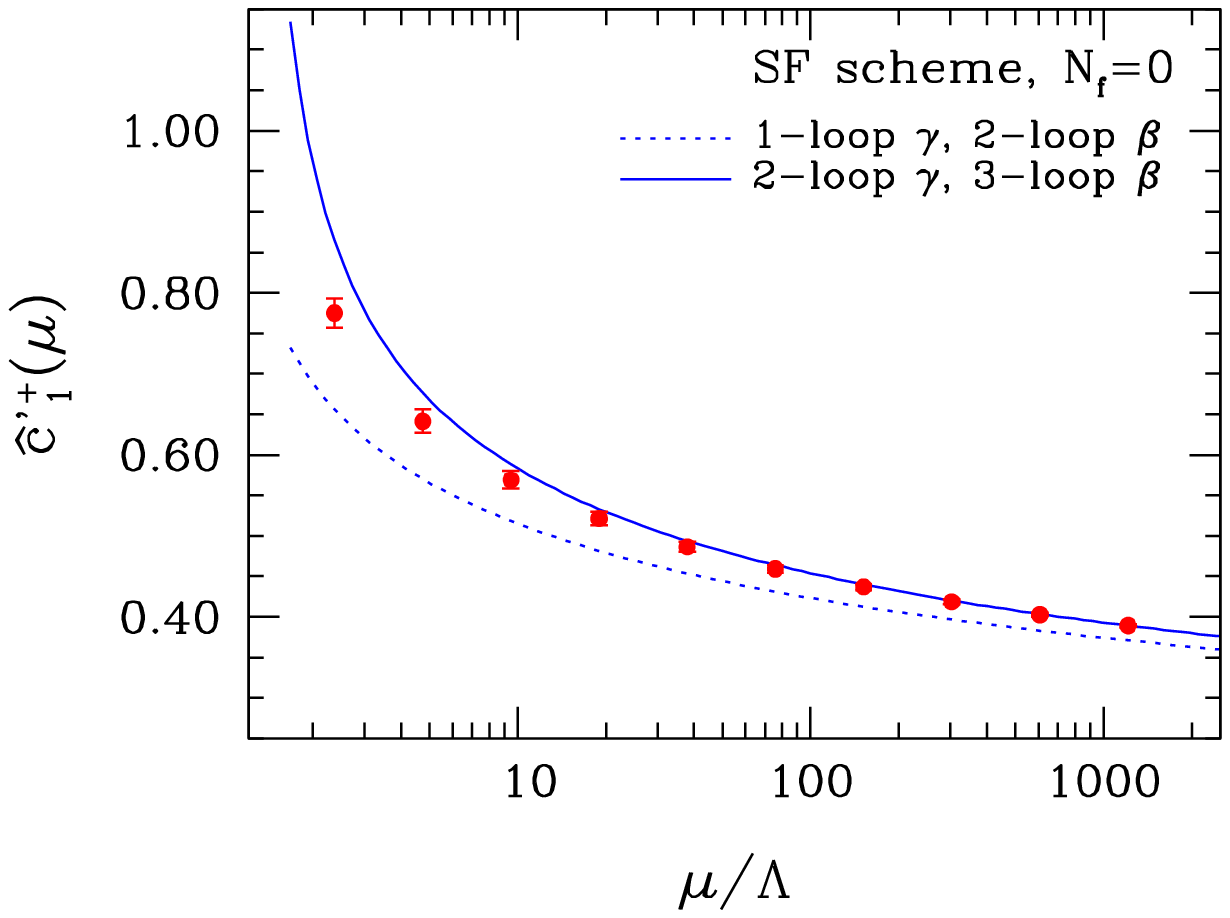}
      \hskip 0.5cm \includegraphics[width=6.8 true cm,angle=0]{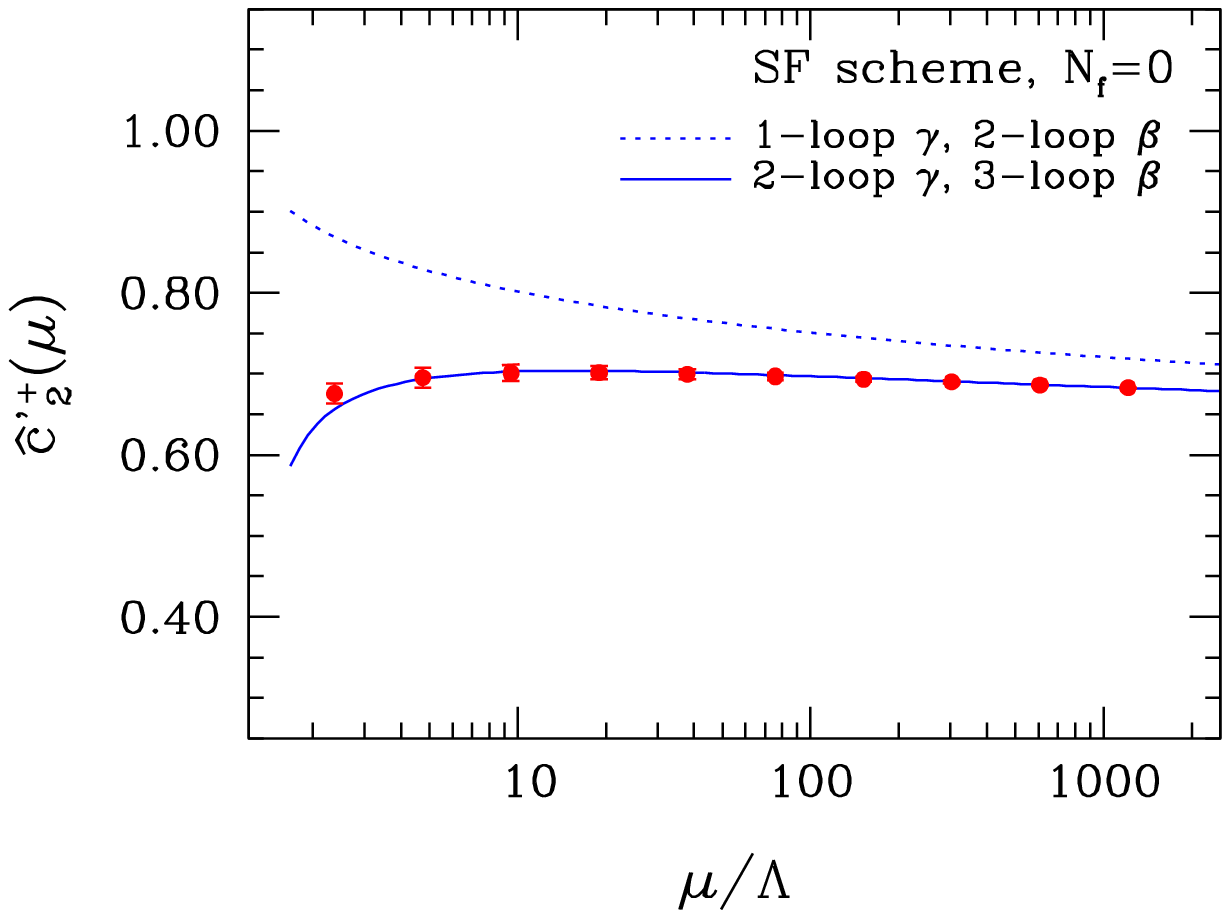}
    \caption{Upper plots: the step scaling functions $\sigma_{12}$ (discrete points) 
      as obtained non-perturbatively. The shaded area is the one sigma band obtained by 
      fitting the points to a polynomial. The dotted (dashed) line is the LO (NLO) perturbative result. 
      Lower plots: RG running of ${\cal Q}'_{1,2}$ obtained non perturbatively (discrete points) 
      at specific values of the renormalization scale $\mu$, in units of $\Lambda$. The lines 
      are perturbative results at the order shown for the Callan-Symanzik $\beta$-function 
      and the operator anomalous dimension $\gamma$.}
    \label{fig:ssf_and_rg}
  \end{center}
\vskip -0.6cm
\end{figure}

In order to compute the RG running of the operators in the
continuum limit as described in section~\ref{RG running}, we need to
fit the results for $\sigma_k(u)$ to some functional form.
We follow the same procedure as for the renormalized
quark mass~\cite{mbar:pap1}, i.e. we adopt the polynomial ansatz
\beq
\sigma_{k}(u)=1 + \sum_{m=1}^M s_m u^m\ ,
\eeq
with $M=2,3,4$ and $s_{1}$ always ($s_2$ possibly) set to its 
perturbative value
\bea
s_1 = {\gamma}_k'^{(0)}\ln 2\ , \qquad\qquad
s_2 = \gamma_k'^{(1)}\ln 2 + \left[
\frac{1}{2}(\gamma_k'^{(0)})^2 + b_0\gamma_k'^{(0)}
\right](\ln 2)^2\ . 
\eea
It is worth mentioning that if $s_2$ is fitted as a free parameter, 
it turns out to lie in the ballpark of perturbation theory. The RG running 
factor ${\hat c}'_k(\mu_{\rm had})={\hat c}'_k(2^n\mu_{\rm had})U'_k(2^n\mu_{\rm had},\mu_{\rm had})$,
which is now a function of the fit parameters only, can be obtained with
a complete control of the systematic effects. We have indeed checked that 
its value is fairly insensitive to the fit ansatz and to whether $s_2$ is 
set to its perturbative value or not. We choose to quote as our 
final results those obtained with $M=3$, $s_1$ fixed by perturbation
theory and $s_2$, $s_3$ kept as free parameters. 
\vskip 0.2cm
In practice, due to constrains imposed by Heavy Quark Spin Symmetry, the number
of independent SF schemes for ${\cal Q}'_k$ is downgraded to four for $k=1$ and to 
eight for $k=2$. These lead to total RGI renormalization factors which are
scheme independent apart from ${\rm O}(a)$ lattice artefacts.
\vskip 0.2cm
The main criterion to define suitable schemes amounts to checking that 
the systematic uncertainty related to truncating the perturbative 
evolution factor ${\hat c}'_k(\mu_{\rm pt})$ of Eq.~(\ref{eq:split}) 
at NLO in the anomalous dimension is well under
control. This in turn requires an estimate of the size of the NNLO
contribution to ${\hat c}'_k(\mu_{\rm pt})$. To this purpose
we have re-computed ${\hat c}'_k(\mu_{\rm pt})$ with two
different values of the NNLO anomalous dimensions $\gamma_k'^{(2)}$:
in the first case we set
$|\gamma_k'^{(2)}/\gamma_k'^{(1)}|=|\gamma_k'^{(1)}/\gamma_k'^{(0)}|$;
in the second case, we guess $\gamma_k'^{(2)}$ by performing a
one-parameter fit to the SSF with $s_1,s_2$ fixed by perturbation
theory, and equating the resulting value of $s_3$ to its perturbative
expression
\bea
s_3 = \gamma_k'^{(2)}\ln 2 &+ \left[
\gamma_k'^{(0)}\gamma_k'^{(1)} + 2b_0\gamma_k'^{(1)}
+ b_1\gamma_k'^{(0)}
\right](\ln 2)^2 + \nn\\ &+ \left[
\frac{1}{6}(\gamma_k'^{(0)})^3 + b_0 (\gamma_k'^{(0)})^2
+\frac{4}{3}b_0^2\gamma_k'^{(0)}
\right](\ln 2)^3 \,.
\eea
For the operator $\cQ'_{1}$, we find that in either case the
central value of the combination
${\hat c}'_k(\mu_{\rm had})\equiv
{\hat c}'_k(\mu_{\rm pt}) U'_k(\mu_{\rm pt},\mu_{\rm had})$
changes by a small fraction of the statistical error, of the order
$0.1$--$0.3$ standard deviations (depending on the renormalization scheme). 
For the operator $\cQ'_2$, which carries 
relatively large NLO anomalous dimensions, the effect can be as large 
as $0.8$--$1.0$ standard deviations. Therefore, we add to ${\hat c}'_2(\mu_{\rm had})$ 
a corresponding systematic uncertainty of this order. 
It has to be stressed that the impact of this extra uncertainty at the 
level of the $B$--$\bar B$ mixing amplitude is not particularly worrying, 
since the matrix element of $\cQ'_2$ enters the latter only at 
${\rm O}(\alpha_s)$, when the static-light theory is matched to QCD.
The results for the SSFs and the operator RG running in the reference schemes 
(see the end of Section~\ref{NPrenSF}) are illustrated in
Figure~\ref{fig:ssf_and_rg}.
\vskip 0.3cm

\section{Matching to hadronic observables}

The RGI operator is connected to its
bare counterpart via the total renormalization factor
${\hat Z}'_{k,\rm RGI}(g_0)$ of Eq.~(\ref{eq:totalren}). We stress
that ${\hat Z}'_{k,\rm RGI}(g_0)$ is a scale-independent quantity, which
moreover depends upon the renormalization scheme only via cutoff
effects. Indeed, it depends on the particular lattice regularization
chosen, though only via the factor ${\hat {\cal Z}}'_k(g_0,a\mu_{\rm
had})$, the computation of which is much less expensive than the
 total RG running factor ${\hat c}'_k(\mu_{\rm had})$.

We have computed ${\cal Z}'_k(g_0,a\mu_{\rm had}),\,
\mu_{\rm{had}}=1/(2L_{\rm max})$ non-perturbatively at four values of
$\beta$ for each scheme and four-fermion operator, and for the four
different static actions under consideration. The total 
renormalization factors are obtained upon multiplying by the corresponding 
running factors on third column of Table~\ref{tab:fit_ztot1}. 
Polynomial interpolations of the form
\beq
  {\hat Z}'_{k,\rm RGI}(g_0)
    =a_{k} +b_{k}(\beta-6) +c_{k}(\beta-6)^2\ ,
  \label{eq:Ztotfit}
\eeq
can be subsequently used to obtain the total renormalization
factor at any value of $\beta$ within the covered range $(6.0 \le \beta \le 6.5)$.
We provide in Table~\ref{tab:fit_ztot1} the resulting fit
coefficients for the HYP2 action in our reference renormalization
schemes. These parametrizations represent our data with an accuracy
of at least $0.3\%$. The contribution from the error in the RG running
factors of Table~\ref{tab:fit_ztot1} has not been included: since these factors
have been computed in the continuum limit, they should be added
in quadrature {\it after} the quantity renormalized with the
factor derived from Eq.~(\ref{eq:Ztotfit}) has been extrapolated itself
to the continuum limit.

\begin{table}[!t]
%\large
\begin{center}
\begin{tabular}{lc@{\hspace{10mm}}clll}
\hline\\[-2.5ex]
$k$ & $s$ & ${\hat c}'_k(\mu_{\rm had})$ & $~~~~a_{k}$ & $~~~~b_{k}$ & $~~~c_{k}$ \\[1.0ex]
\hline\\[-2.5ex]
1     & 1 & 0.777(17) & 0.5731(11) & -0.171(11) & 0.082(25)\\
2     & 3 & 0.675(12) & 0.7258(14) & -0.061(14) & 0.016(33) \\[1.0ex]
\hline\\[-2.5ex]
\end{tabular}
\caption{ Running ${\hat c}'_k(\mu_{\rm had})$ (with $\mu_{\rm
  had}=(2L_{\rm max})^{-1}$) and fits to the total renormalization factor 
of Eq.~\protect\ref{eq:totalren}. }
\label{tab:fit_ztot1}
\end{center}
\end{table}

\section{SF correlators for the bare matrix elements}

In order to simulate the physical matrix elements needed for the $B$-parameter, we adopt a 
formalism similar to the one described in the previous sections, where the heavy quark field 
is described by the HYP2 static action, while the light quark field is described by the tmQCD 
action including the Sheikoleslami-Wohlert term with non-perturbatively defined $c_{\rm sw}$.  
The interpolating operators of the $B_s$- and $\bar B_s$-mesons are provided by the boundary 
sources $\Sigma_{sh}$ and $\Sigma'_{s\bar h}$. Correlation functions are then constructed by 
inserting a bilinear or a four-fermion operator in the bulk of the SF. Accordingly, the 
building blocks of the computation are given by
\begin{equation}
\label{twopoint}
f_{\rm\scriptscriptstyle X}(x_0) = -\frac{\ a^3}{\ 2\ }\sum_{\bx}\langle X_{hs}(x)\Sigma_{sh}\rangle \ , \qquad
f'_{\rm\scriptscriptstyle X}(x_0) = -\frac{\ a^3}{\ 2\ }\sum_{\bx}\langle \Sigma'_{s\bar h}X_{\bar hs}(x)\rangle \ ,
\end{equation}
\begin{equation}
\label{threepoint}
F_{\rm\scriptscriptstyle Y}(x_0) = a^3\sum_{\bx}\langle \Sigma'_{s\bar h}Y_{hs\bar h s}(x)\Sigma_{sh}\rangle\ ,
\end{equation}
where $X=A_0^{\rm stat},V_0^{\rm stat}$ and $Y=\cQ'_{1,2}$. To be precise, the 
extraction of the $B$-parameter of \Eq{Bpars} requires that the matrix elements 
of $\cQ'_{1,2}$ be normalized by the square of the decay matrix element of the
$B_s$-meson mediated by the static axial current.  Since the latter is rotated at 
full twist into a linear combination of axial and vector currents, we obtain the single 
contributions to the $B$-parameter from the plateau region of the ratios
\begin{equation}
R_i(x_0) = \frac{3}{8}\frac{F_{\scriptscriptstyle \cQ'_i}(x_0)}{[2h_{\rm\scriptscriptstyle A-iV}(x_0)][2h'_{\rm\scriptscriptstyle A-iV}(T-x_0)]}\ , \qquad i=1,2 \ ,
\end{equation}
where
% \begin{align}
% h_{\rm\scriptscriptstyle A-iV}(x_0) & = \frac{1}{\sqrt{2}}\left[Z^{\rm stat}_{\rm\scriptscriptstyle A,RGI}f_{\rm\scriptscriptstyle A}^{\rm stat}(x_0) - Z^{\rm stat}_{\rm\scriptscriptstyle V,RGI}f_{\rm\scriptscriptstyle V}^{\rm stat}(x_0)\right]\ , \nonumber \\
% h'_{\rm\scriptscriptstyle A-iV}(x_0) & = \frac{1}{\sqrt{2}}\left[Z^{\rm stat}_{\rm\scriptscriptstyle A,RGI}f'_{\rm\scriptscriptstyle A}(x_0) - Z^{\rm stat}_{\rm\scriptscriptstyle V,RGI}f'_{\rm\scriptscriptstyle V}(x_0)\right]\ .
% \end{align}
\begin{equation}
h_{\rm\scriptscriptstyle A-iV}(x_0)  = \frac{1}{\sqrt{2}}\left[Z^{\rm stat}_{\rm\scriptscriptstyle A,RGI}f_{\rm\scriptscriptstyle A}^{\rm stat}(x_0) - Z^{\rm stat}_{\rm\scriptscriptstyle V,RGI}f_{\rm\scriptscriptstyle V}^{\rm stat}(x_0)\right]\ .
\end{equation}
The RGI axial constant $Z^{\rm stat}_{\rm\scriptscriptstyle A,RGI}$ 
has been non-per\-tur\-ba\-ti\-ve\-ly computed in \cite{Heitger:2003xg,Della Morte:2005yc}. 
The scale independent ratio 
$Z^{\rm stat}_{\rm\scriptscriptstyle V,RGI}/Z^{\rm stat}_{\rm\scriptscriptstyle A,RGI}$ 
is taken from \cite{Palombi:2007dt}. We performed simulations
at $\beta=6.0,6.1,6.2$ with the strange quark mass set to physical values 
as in \cite{Rolf:2002gu}. Lattice parameters are collected in Table~\ref{tab:parameters}.

\section{Analysis of the excited state contaminations}

\begin{figure}[!ht]
  \begin{center}
    \begin{minipage}{.45\linewidth}
      \hskip -0.7cm\includegraphics[width=6.5 true cm,angle=-90]{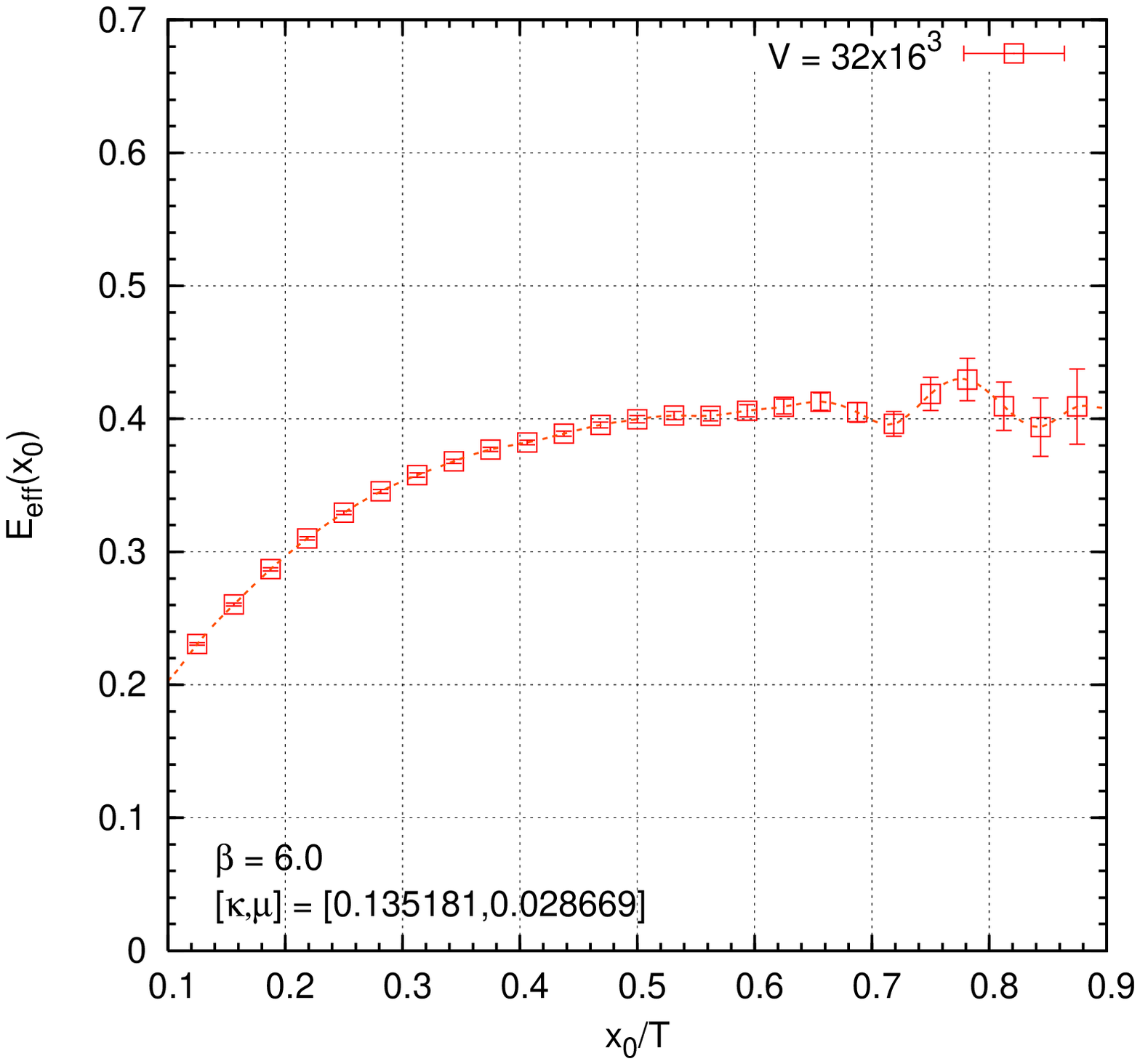}
    \end{minipage}
    \begin{minipage}{.45\linewidth}
      \includegraphics[width=6.5 true cm,angle=-90]{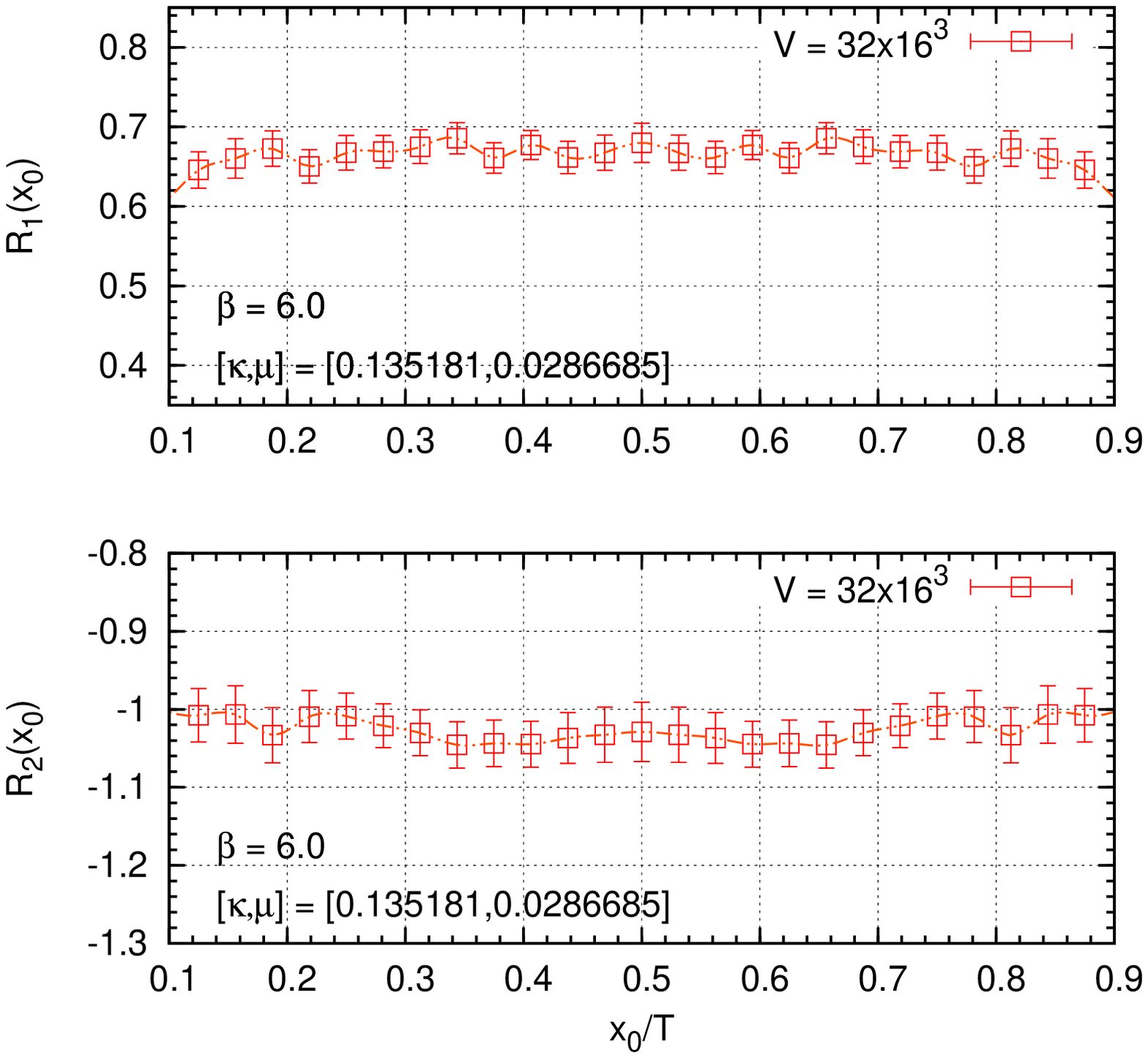}
    \end{minipage}
    \vskip 0.0cm
    \caption{The binding energy and the contributions to the $B_s$-parameter at $\beta=6.0$.}
    \label{fig:fig1}
  \end{center}
\vskip -0.6cm
\end{figure}

\begin{figure}[!ht]
  \begin{center}
    \begin{minipage}{.45\linewidth}
      \hskip -0.7cm\includegraphics[width=6.5 true cm,angle=-90]{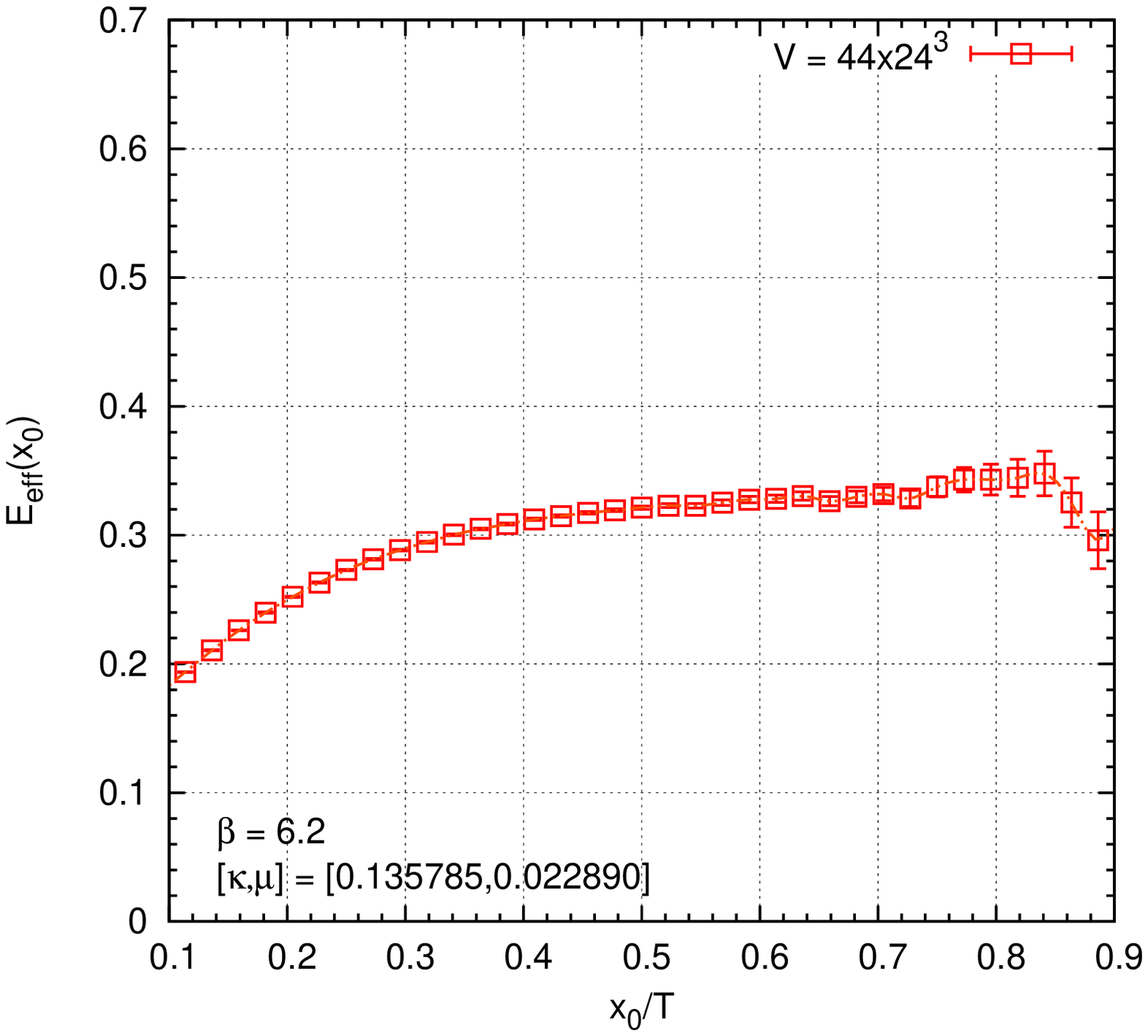}
    \end{minipage}
    \begin{minipage}{.45\linewidth}
      \includegraphics[width=6.5 true cm,angle=-90]{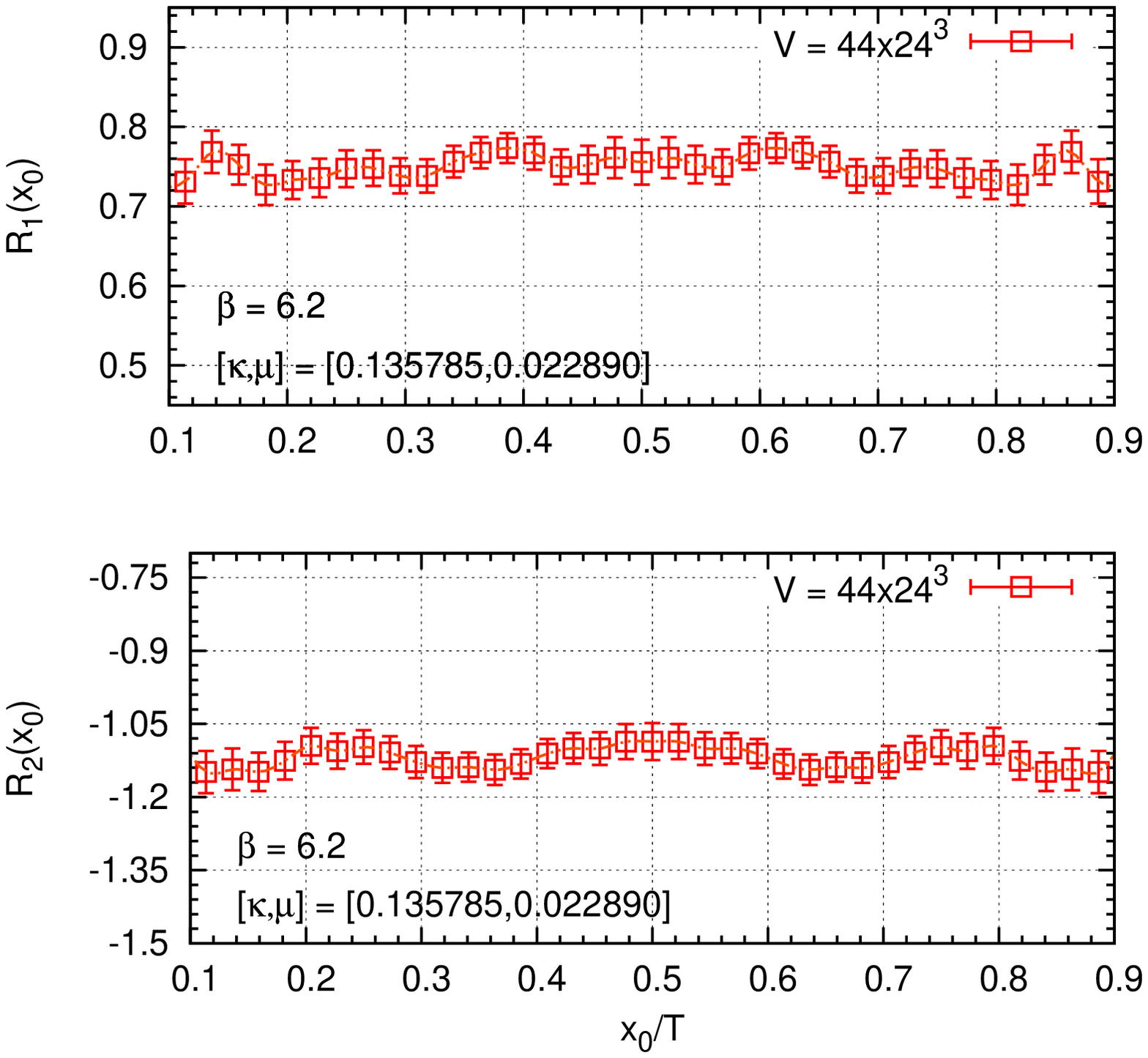}
    \end{minipage}
    \vskip 0.0cm
    \caption{The binding energy and the contributions to the $B_s$-parameter at $\beta=6.2$.}
    \label{fig:fig2}
  \end{center}
\vskip -0.6cm
\end{figure}

The standard way to identify a plateau interval for a three-point correlation
function such as \Eq{threepoint} is to analyse the exponential decay rate of 
the corresponding meson propagator $h_{\rm\scriptscriptstyle A-iV}$, obtained 
via the binding energy
\vskip 0.4cm
\TABLE[!hr]{
  \scriptsize
  \centering
  \vbox{\vskip 0.0cm}
  \begin{tabular}{|c|c|c|c|c|}
    \hline
    $\beta$ & $T\times L^3$ & $\kappa_{\rm cr}$ & $\kappa$ & $\mu$ \\
    \hline
    $6.0$ & $32\times 16^3$ & $0.135196$ & $0.135181$ & $0.028669$ \\
    $6.1$ & $38\times 24^3$ & $0.135665$ & $0.135650$ & $0.028532$ \\
    $6.2$ & $44\times 24^3$ & $0.135795$ & $0.135785$ & $0.022890$ \\
    \hline
  \end{tabular}
  \caption{Lattice parameters\label{tab:parameters}}
}
\vskip -0.6cm
\begin{equation}
aE_{\rm eff}(x_0) = \frac{1}{2}\log\left\{\frac{h_{\rm\scriptscriptstyle A-iV}(x_0-a)}
{h_{\rm\scriptscriptstyle A-iV}(x_0+a)}\right\}\ . 
\end{equation}
This procedure may work only provided that the lowest value 
$x_0^{\rm min}$, at which the fundamental state is numerically 
isolated, fulfills the condition $x_0^{\rm min}<T/2$. 
Correspondingly, the interval $[x_0^{\rm min},T-x_0^{\rm min}]$ 
can be certainly used to extract the plateau value of the 
three-point correlator. Unfortunately, this is not the case, 
as shown in Figs.~\ref{fig:fig1} and \ref{fig:fig2} (left): due to the small mass gap between 
the lowest and the first excited states in the static-light 
channel, the plateau starts at about the middle of the lattice. 
Simulations at larger time extensions are increasingly expensive
owing to the exponential rise of the noise-to-signal ratio 
related to static propagators.
Irrespective of this, the observables $R_i$ are characterized by 
a very flat time dependence; examples are provided 
by Figs.~\ref{fig:fig1} and \ref{fig:fig2} (right). In order to understand this  
behaviour, we perform an expansion of Eqs.~(\ref{twopoint},\ref{threepoint}) 
through the insertion of complete sets of Hamiltonian eigenstates. Assuming 
that the excited contributions in the vacuum channel may be disregarded, one 
easily arrives at the representation
\begin{equation}
\label{Bparexp}
R_i(x_0) = B_i^{(0,0)}\frac{1 + \sum_{(n,m)\ne(0,0)}^\infty \frac{B_i^{(n,m)}}{B_i^{(0,0)}}f_{nm}g_{nm}
{\rm e}^{-(T-x_0)\Delta^{(B)}_{n0}}{\rm e}^{-x_0\Delta^{(B)}_{n0}}}
{1 + \sum_{(n,m)\ne(0,0)}^\infty \ \ \ \ \ \ \ f_{nm}g_{nm}
{\rm e}^{-(T-x_0)\Delta^{(B)}_{n0}}{\rm e}^{-x_0\Delta^{(B)}_{n0}}}\ , 
\end{equation}
where $\Delta^{(B)}_{n0} = E^{(B)}_n - E^{(B)}_0$ is the energy gap 
between the $n$-th Hamiltonian eigenstate $|n,B\rangle$ with the 
quantum numbers of a static $B_s$-meson and the fundamental state. 
Moreover, 
\begin{equation}
B_i^{(n,m)} = \frac{\langle n,B|\cQ'_i|m,B\rangle}{\frac{8}{3}\langle n,B|A_0|0,0\rangle\langle 0,0|A_0|m,B\rangle}\ ,
\end{equation}
\begin{equation}
\label{melements}
f_{nm} = \frac{\langle \ri_B|n,B\rangle\langle m,B|\ri_B\rangle}{\langle \ri_B|0,B\rangle\langle 0,B|\ri_B\rangle}\ , \qquad
g_{nm} = \frac{\langle n,B|A_0|0,0\rangle\langle 0,0|A_0|m,B\rangle}{\langle 0,B|A_0|0,0\rangle\langle 0,0|A_0|0,B\rangle}\ .
\end{equation}
Here $|\ri_B\rangle$ represents the SF boundary state corresponding 
to the action of the bilinear sources \Eq{bilinearsource} on the 
vacuum. In particular, one should observe that $B_i^{(n,m)}$ 
represents a generalization of the $B$-parameter, describing the 
particle-antiparticle mixing of excited states. Numerator and 
denominator of \Eq{Bparexp} look quite similar. They only differ by 
the weighting coefficients $z_{nm} = B_i^{(n,m)}/B_i^{(0,0)}$. 

The hypothetical condition $z_{nm}\approx 1$ would act on the ratios $R_i(x_0)$ 
as an additional damping factor of the excited state contaminations, together 
with the exponential decays due to the mass gaps $\Delta_{nm}^{(B)}$.  In 
practice, what really matters for our concern is the first excited contribution, 
because already the second excitation is reasonably expected to compete with the 
$0^{++}$ glueball ($r_0m_G\approx 1.7\ {\rm GeV}$). Data suggest that $z_{10}$ could be
quite close to one. 

Since the quantum states $|B,n\rangle$ and $|B,0\rangle$ differ only by their mass, we are
led to speculate about the mass dependence of the generalized $B$-parameters $B^{(n,m)}_i$.
Although quantitative statements are highly non-trivial, it is not difficult to identify at 
least one extreme situation where the limits $z_{nm}=1$ could be realized. This is a scenario 
in which $B_i^{(n,m)}$ is weakly dependent upon the mass of the external states. 
If $B_i^{(n,m)}$ is close to $B_i^{(0,0)}$, then their ratio will be close to one.
This picture imposes no restrictions on the value of $B_i^{(0,0)}$. 

An apparently different possibility is represented by the VSA, which implies $B_i^{(n,m)}=1$, and
consequently $z_{nm}=1$. Though speculative, it is not unreasonable that the violation of the VSA 
depends weakly upon the mass of the external states and is responsible in the end
for the realization of the above-mentioned scenario. 

A quantitative check of the suppression of excited state contaminations in 
$R_{1,2}(x_0)$ would be provided by the level of stability of the observed plateaux 
under a variation of the boundary interpolating operators. In the framework of the
Schr\"odinger functional this possibility is explored via the introduction
of boundary wave functions like in \cite{DellaMorte:2003mn}. Unfortunately,
the computational price required at present for the practical implementation of this
smearing technique amounts to giving up one of the boundary summations of 
Eq.~(\ref{bilinearsource}), with a corresponding increase of the statistical
error by a factor of $\sqrt{L^3}$. The situation is even worse with a three-point
correlator such as Eq.~(\ref{threepoint}), which has interpolating sources on
both boundaries. In this case the introduction of smearing wave functions increases 
the statistical noise by a factor of $L^3$, which makes the check useless. This 
problem can be hopefully overcome through the implementation of a SF all-to-all 
propagator like proposed in \cite{Giusti:2004yp,Foley:2005ac,Luscher:2007se}. This is 
currently under way.

\section{A two-state stochastic model}

\begin{figure}[!t]
  \begin{center}
    \begin{minipage}{.45\linewidth}
      \hskip -0.7cm\includegraphics[width=6.5 true cm,angle=-90]{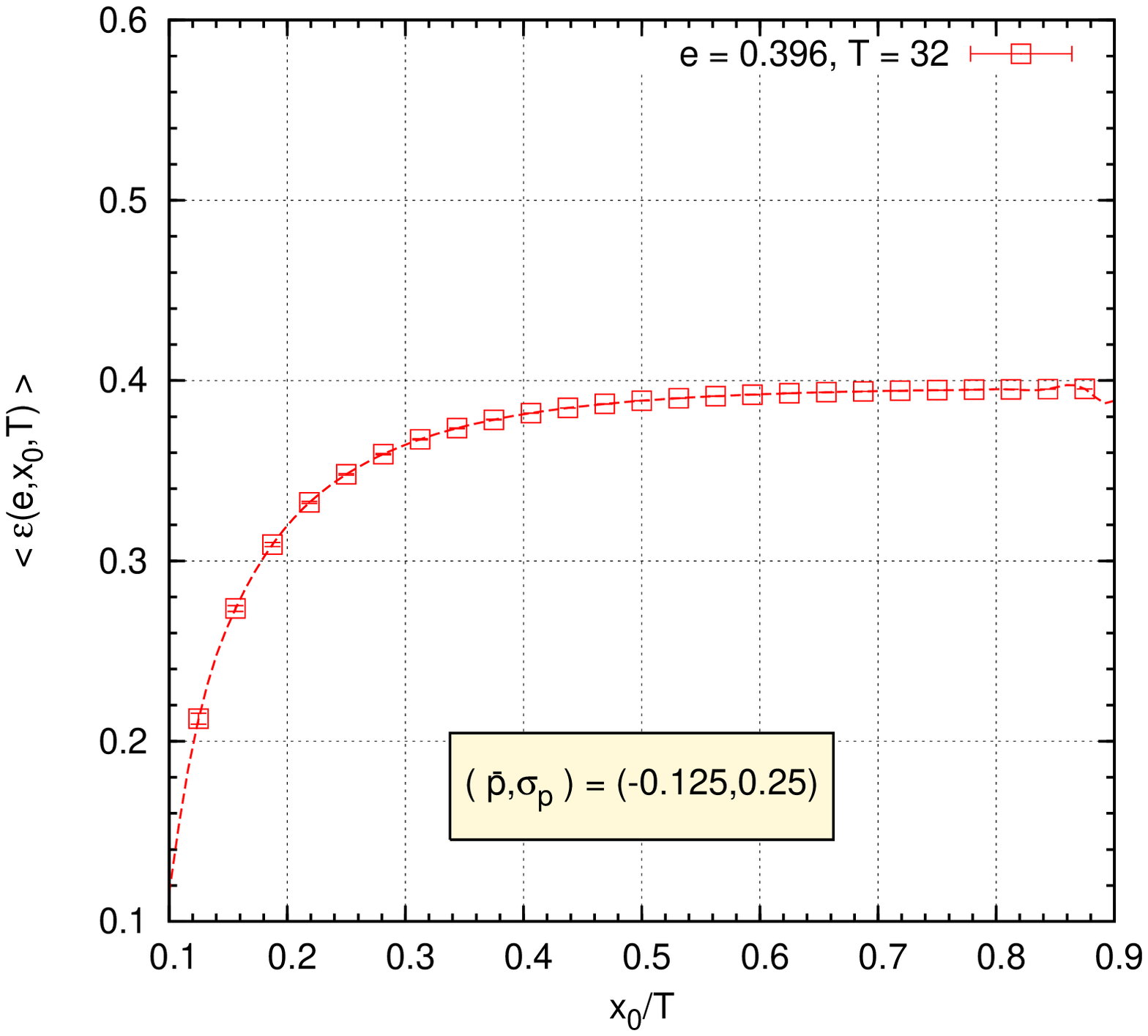}
    \end{minipage}
    \begin{minipage}{.45\linewidth}
      \includegraphics[width=6.5 true cm,angle=-90]{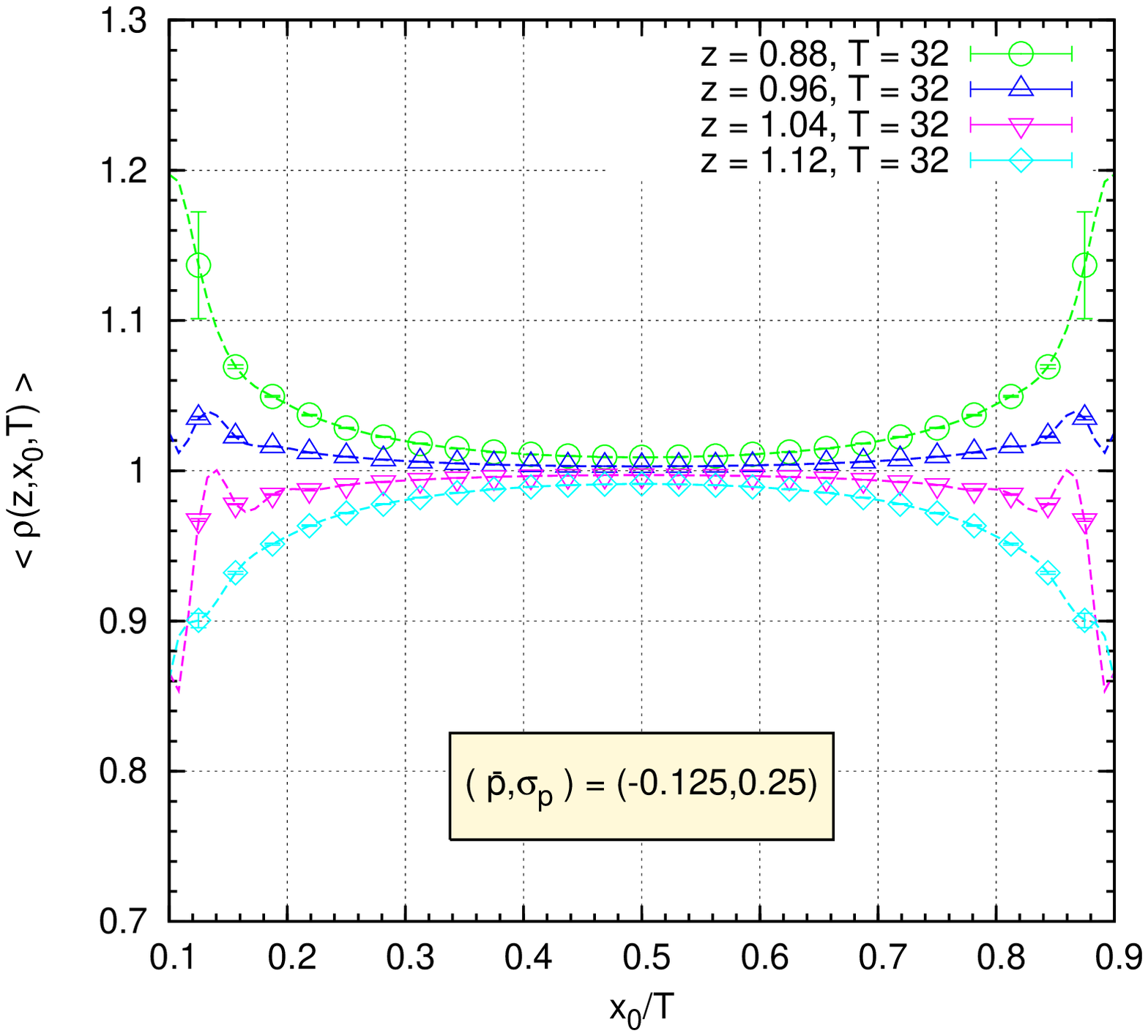}
    \end{minipage}
    \vskip 0.0cm
    \caption{The binding energy and the contributions to the $B_s$-parameter within a two-state stochastic model.}
    \label{fig:fig3}
  \end{center}
\vskip -0.6cm
\end{figure}

In order to have a qualitative view about the impact of large deviations of $z_{10}$ 
from one on the time dependence of $R_i(x_0)$, we consider a two-state model. Here, 
the binding energy and the contributions to the $B$-parameters are described 
by the stochastic variables
\begin{align}
\epsilon(e,x_0,T) & = e + \frac{1}{2}\log\left\{\frac{1-p{\rm e}^{-\Delta(x_0-1)}}{1-p{\rm e}^{-\Delta(x_0+1)}}\right\}\ , \\[1.2ex]
\rho(z,x_0,T) & = \frac{1-zp\left[{\rm e}^{-\Delta(x_0)} + {\rm e}^{-\Delta(T-x_0)}\right]}{1-p\left[{\rm e}^{-\Delta(x_0)}
+ {\rm e}^{-\Delta(T-x_0)}\right]}\ ,
\end{align}
where $p$ and $\Delta$ are differently distributed random coefficients, 
while $z$ and $e$ parametrize $z_{10}$ and $E_0^{(B)}$. Obviously, $\Delta$
is meant to represent the energy gap of the first excited state. From a 
two-state analysis of $E_{\rm eff}(x_0)$, it is roughly known that 
$a\Delta\approx 0.22(3)$. Therefore, we model this variable according to a 
Gaussian distribution probability, i.e.
\begin{equation}
P(\Delta) = \frac{1}{\sigma_\Delta\sqrt{2\pi}}{\rm exp}\left(-\frac{\Delta-\bar\Delta}{2\sigma_{\Delta}^2}\right)\ , 
\qquad (\bar\Delta,\sigma_\Delta) = (0.22,0.03)\ .
\end{equation}
On the other hand, $p$ is supposed to represent the product of the matrix 
elements $f_{10}$ and $g_{10}$, defined in Eq.~(\ref{melements}). Choosing a distribution probability for this variable
is delicate, because we are largely ignorant about the projection of the SF 
boundary state $|\ri_B\rangle$ onto the first excited state $|1,B\rangle$ and
the decay constant of the latter. We can heuristically expect that
\begin{equation}
h_{10} = \frac{\langle 1,B|A_0|0,0\rangle}{\langle 0,B|A_0|0,0\rangle} = 
\frac{m_B^*f_B^*}{m_Bf_B} \approx 1 \ . 
\end{equation}
Nevertheless, if we believe that $|\ri_B\rangle$ is well projected onto 
$|0,B\rangle$, then $f_{10}\approx 0$. In this case we should choose
a probability distribution of $p$ peaked around $p=0$. By contrast, if we
believe that $|\ri_B\rangle$ is a balanced mixture of $|0,B\rangle$ and
$|1,B\rangle$, it follows that $f_{10}\approx 1$. It makes sense 
to assume a given sign for $p$ and not to allow for fluctuations of the opposite
sign. A flexible distribution probability allowing for a definite sign 
is the Log-normal distribution, defined by
\begin{equation}
P(p;\bar p,\sigma_p) = \frac{1}{p\sigma_p\sqrt{2\pi}}\exp\left\{-\frac{(\ln p - \bar p)^2}{2\sigma_p^2}\right\}\ . 
\end{equation}
Having produced $N$ samples $\{\Delta_i\}_{i=1\dots N}$ and $\{p_i\}_{i=1\dots N}$
of $\Delta$ and $p$, we approximate the ensemble averages of $\mu(z,x_0,T)$ and $\epsilon(e,x_0,T)$ via
\begin{align}
{\cal E}(e,x_0,T) & = \langle\epsilon(e,x_0,T)\rangle \simeq e + \frac{1}{2N}\sum_{i=1}^N\log\left\{\frac{1-p_i{\rm e}^{-\Delta_i(x_0-1)}}{1-p{\rm e}^{-\Delta_i(x_0+1)}}\right\}\ , \\[1.2ex]
{\cal R}(z,x_0,T) & = \langle\rho(z,x_0,T)\rangle \simeq \frac{1}{N}\sum_{i=1}^N \frac{1-zp_i\left[{\rm e}^{-\Delta_i(x_0)} + {\rm e}^{-\Delta_i(T-x_0)}\right]}{1-p_i\left[{\rm e}^{-\Delta_i(x_0)}
+ {\rm e}^{-\Delta_i(T-x_0)}\right]}\ .
\end{align}
The ensemble averages $\cal E$ and $\cal R$ are now functions of the distribution
parameters $\bar p$ and $\sigma_p$, which can be varied in order to change the
shape of the distribution. One of the worst cases we considered is the one corresponding
to $(\bar p = -1/8, \sigma_p=1/4)$,  i.e a Log-normal distribution
peaked around $\exp(-\bar p)\simeq 0.88$. In the spirit of the two-state model, such
a distribution describes a large overlap between the interpolating boundary state and 
the first excited one. As shown in Fig.~\ref{fig:fig3}, the binding energy resembles
very closely the one of Fig.~\ref{fig:fig1}. The shapes of the $B$-parameter corresponding
of different choices of $z$ suggest that $z_{10}$ could be very close to one in the real case,
thus supporting our interpretation in terms of the VSA. 

\section{Conclusions}

$B^0-\bar{B}^0$ mixing remains among the most important processes that
are required to pin down the elements of the CKM matrix precisely. 
However, in order to constrain the unitarity triangle sufficiently well 
and to look for signs of new physics, theoretical uncertainties associated 
with hadronic effects must be reduced. In this talk we have 
reported on a new strategy for the computation of the heavy-light 
$B$-parameters in lattice QCD, based on tmQCD and HQET. Its main advantage 
is the exact absence of mixing under renormalization, which plagues standard Wilson 
fermions, at the same computational cost of a Wilson-type regularization. 
We have described our fully non-perturbative calculation
of the relations between parity-odd, static-light four-quark operators
in quenched lattice QCD and their renormalized counterparts. We have
also described our first experiences with the computation of the bare 
matrix elements for the $B_s-\bar B_s$ mixing and their excited state 
contaminations. Before attempting a continuum extrapolation of the matrix 
elements, a deeper analysis of the excited state contributions has to be 
done and we hope that the all-to-all propagator, like proposed in 
\cite{Giusti:2004yp,Foley:2005ac,Luscher:2007se} will be of great help there. 


\begin{thebibliography}{99}

%\cite{Frezzotti:2000nk}
\bibitem{Frezzotti:2000nk}
  R.~Frezzotti, P.~A.~Grassi, S.~Sint and P.~Weisz,
  %``Lattice QCD with a chirally twisted mass term,''
  JHEP {\bf 0108} (2001) 058
  [arXiv:hep-lat/0101001]. \\[0.5ex]
  %%CITATION = JHEPA,0108,058.%%

%\cite{Buchalla:1996ys}
\bibitem{Buchalla:1996ys}
  G.~Buchalla,
  %``Renormalization of Delta(B) = 2 transitions in the static limit beyond
  %leading logarithms,''
  Phys.\ Lett.\  B {\bf 395} (1997) 364
  [arXiv:hep-ph/9608232].
  %%CITATION = PHLTA,B395,364;%%

%\cite{Palombi:2006pu}
\bibitem{Palombi:2006pu}
  F.~Palombi, M.~Papinutto, C.~Pena and H.~Wittig,
  %``A strategy for implementing non-perturbative renormalization of
  %heavy-light four-quark operators in the static approximation,''
  JHEP {\bf 0608} (2006) 017
  [arXiv:hep-lat/0604014].

%\cite{Palombi:2007dr}
\bibitem{Palombi:2007dr}
  %%CITATION = JHEPA,0608,017;%%
  F.~Palombi, M.~Papinutto, C.~Pena and H.~Wittig,
  %``Non-perturbative renormalization of static-light four-fermion operators in
  %quenched lattice QCD,''
  arXiv:0706.4153 [hep-lat].
  %%CITATION = ARXIV:0706.4153;%%

%\cite{Luscher:1992an}
\bibitem{Luscher:1992an}
  M.~L\"uscher, R.~Narayanan, P.~Weisz and U.~Wolff,
  %``The Schrodinger functional: A Renormalizable probe for nonAbelian gauge
  %theories,''
  Nucl.\ Phys.\  B {\bf 384} (1992) 168
  [arXiv:hep-lat/9207009].
  %%CITATION = NUPHA,B384,168;%%

%\cite{Pena:lat07}
\bibitem{Pena:lat07}
  P.~Dimopoulos, G.~Herdoiza, F.~Palombi, M.~Papinutto, C.~Pena, A.~Vladikas and H.~Wittig,
  %``Non-perturbative renormalization of static-light four-fermion operators in
  %quenched lattice QCD,''
  PoS(LAT2007)368 arXiv:0710.2862 [hep-lat].
  %%CITATION = ARXIV:0710.2862;%%

%\cite{Luscher:1996sc}
\bibitem{Luscher:1996sc}
  M.~Luscher, S.~Sint, R.~Sommer and P.~Weisz,
  %``Chiral symmetry and O(a) improvement in lattice QCD,''
  Nucl.\ Phys.\  B {\bf 478} (1996) 365
  [arXiv:hep-lat/9605038].
  %%CITATION = NUPHA,B478,365;%%

%\cite{Eichten:1989zv}
\bibitem{Eichten:1989zv}
  E.~Eichten and B.~R.~Hill,
  %``An Effective Field Theory for the Calculation of Matrix Elements Involving
  %Heavy Quarks,''
  Phys.\ Lett.\  B {\bf 234} (1990) 511.
  %%CITATION = PHLTA,B234,511;%%

%\cite{Della Morte:2005yc}
\bibitem{Della Morte:2005yc}
  M.~Della Morte, A.~Shindler and R.~Sommer,
  %``On lattice actions for static quarks,''
  JHEP {\bf 0508} (2005) 051
  [arXiv:hep-lat/0506008].
  %%CITATION = JHEPA,0508,051;%%

%\cite{Necco:2001gh}
\bibitem{Necco:2001gh}
  S.~Necco and R.~Sommer,
  %``Testing perturbation theory on the N(f) = 0 static quark potential,''
  Phys.\ Lett.\  B {\bf 523} (2001) 135
  [arXiv:hep-ph/0109093].
  %%CITATION = PHLTA,B523,135;%%


%\cite{Bode:1999sm}
\bibitem{Bode:1999sm}
  A.~Bode, P.~Weisz and U.~Wolff,
  %``Two loop computation of the Schroedinger functional in lattice QCD,''
  Nucl.\ Phys.\  B {\bf 576} (2000) 517
  [Erratum-ibid.\  B {\bf 600} (2001\ ERRAT,B608,481.2001) 453]
  [arXiv:hep-lat/9911018].
  %%CITATION = NUPHA,B576,517;%%

%\cite{mbar:pap1}
\bibitem{mbar:pap1}
S.~Capitani, M.~L\"uscher, R.~Sommer and H.~Wittig,
%``Non-perturbative quark mass renormalization in quenched lattice QCD,''
Nucl.\ Phys.\ B {\bf 544} (1999) 669
[arXiv:hep-lat/9810063].
%%CITATION = HEP-LAT 9810063;%%

%\cite{Hasenfratz:2001hp}
\bibitem{Hasenfratz:2001hp}
  A.~Hasenfratz and F.~Knechtli,
  %``Flavor symmetry and the static potential with hypercubic blocking,''
  Phys.\ Rev.\  D {\bf 64} (2001) 034504
  [arXiv:hep-lat/0103029].
  %%CITATION = PHRVA,D64,034504;%%

%\cite{Heitger:2003xg}
\bibitem{Heitger:2003xg}
  J.~Heitger, M.~Kurth and R.~Sommer,
  %``Non-perturbative renormalization of the static axial current in  quenched
  %QCD,''
  Nucl.\ Phys.\  B {\bf 669} (2003) 173
  [arXiv:hep-lat/0302019].
  %%CITATION = NUPHA,B669,173;%%

%\cite{Palombi:2007dt}
\bibitem{Palombi:2007dt}
  F.~Palombi,
  %``Non-perturbative renormalization of the static vector current and its
  %O(a)-improvement in quenched QCD,''
  arXiv:0706.2460 [hep-lat].
  %%CITATION = ARXIV:0706.2460;%%

%\cite{Rolf:2002gu}
\bibitem{Rolf:2002gu}
  J.~Rolf and S.~Sint,
  %``A precise determination of the charm quark's mass in quenched QCD,''
  JHEP {\bf 0212} (2002) 007
  [arXiv:hep-ph/0209255].
  %%CITATION = JHEPA,0212,007;%%

%\cite{DellaMorte:2003mn}
\bibitem{DellaMorte:2003mn}
  M.~Della Morte, S.~D\"urr, J.~Heitger, H.~Molke, J.~Rolf, A.~Shindler and R.~Sommer,
  %``Lattice HQET with exponentially improved statistical precision,''
  Phys.\ Lett.\  B {\bf 581} (2004) 93
  [Erratum-ibid.\  B {\bf 612} (2005) 313]
  [arXiv:hep-lat/0307021].
  %%CITATION = PHLTA,B581,93;%%

%\cite{Giusti:2004yp}
\bibitem{Giusti:2004yp}
  L.~Giusti, P.~Hernandez, M.~Laine, P.~Weisz and H.~Wittig,
  %``Low-energy couplings of QCD from current correlators near the chiral
  %limit,''
  JHEP {\bf 0404} (2004) 013
  [arXiv:hep-lat/0402002].
  %%CITATION = JHEPA,0404,013;%%

%\cite{Foley:2005ac}
\bibitem{Foley:2005ac}
  J.~Foley, K.~Jimmy Juge, A.~O'Cais, M.~Peardon, S.~M.~Ryan and J.~I.~Skullerud,
  %``Practical all-to-all propagators for lattice QCD,''
  Comput.\ Phys.\ Commun.\  {\bf 172} (2005) 145
  [arXiv:hep-lat/0505023].
  %%CITATION = CPHCB,172,145;%%

%\cite{Luscher:2007se}
\bibitem{Luscher:2007se}
  M.~L\"uscher,
  %``Local coherence and deflation of the low quark modes in lattice QCD,''
  JHEP {\bf 0707} (2007) 081
  arXiv:0706.2298 [hep-lat].
  %%CITATION = JHEPA,0707,081;%%

\end{thebibliography}
\end{document}